\documentclass[
superscriptaddress,
amsmath,amssymb,aps,prb,twocolumn,
floatfix
]{revtex4-2}
\usepackage{graphicx}
\usepackage{dcolumn}
\usepackage{placeins}
\usepackage{bm}
\usepackage{float}
\usepackage{xcolor}
\usepackage{comment}
\usepackage[normalem]{ulem}
\usepackage[USenglish,american]{babel}
\usepackage[colorlinks,bookmarks=false,citecolor=blue,linkcolor=blue,urlcolor=blue]{hyperref}
\usepackage{graphicx}
\usepackage{latexsym}
\usepackage{epstopdf}
\usepackage{amsfonts}
\usepackage{subfigure}
\usepackage{float}

\expandafter\ifx\csname package@font\endcsname\relax\else
\expandafter\expandafter
\expandafter\usepackage
\expandafter\expandafter
\expandafter{\csname package@font\endcsname}
\fi
\begin{document}
\title{From hard-core to soft-core bosons in a shell trap}

\author{Gabriele Costa}
\email{gabriele.costa@studenti.unime.it}
\affiliation{Dipartimento di Scienze Matematiche e Informatiche, Scienze Fisiche e Scienze della Terra, Università degli Studi di Messina, Viale F. Stagno d’Alcontres 31, 98166 Messina, Italy}
\affiliation{Institute for Chemical-Physical Processes, National Research Council of Italy (IPCF-CNR),
98158 Messina (Italy)}

\author{Santi Prestipino}
\email{sprestipino@unime.it}
\affiliation{Dipartimento di Scienze Matematiche e Informatiche, Scienze Fisiche e Scienze della Terra, Università degli Studi di Messina, Viale F. Stagno d’Alcontres 31, 98166 Messina, Italy}
\affiliation{Institute for Chemical-Physical Processes, National Research Council of Italy (IPCF-CNR),
98158 Messina (Italy)}

\author{Fabio Cinti}
\email{fabio.cinti@unifi.it}
\affiliation{Dipartimento di Fisica e Astronomia, Universit\`a di Firenze, I-50019, Sesto Fiorentino (FI), Italy}
\affiliation{INFN, Sezione di Firenze, I-50019, Sesto Fiorentino (FI), Italy}
               
\author{Matteo Ciardi}
\email{matteo.ciardi@tuwien.ac.at}
\affiliation{Institute for Theoretical Physics, TU Wien, Wiedner Hauptstraße 8-10/136, 1040 Vienna, Austria}

\begin{abstract}
We study the low-temperature states of a system of spinless bosons trapped in a broad spherical shell.
In our virtual (path-integral Monte Carlo) experiments, we specifically examine how the equilibrium structure and overall superfluidity of the system are affected by a change of the interparticle-potential core from hard to increasingly soft.
In this manner we bridge the gap between the behaviors of two species of bosons: Helium-like atoms, which self-assemble into ordinary crystalline solids, and Rydberg-dressed atoms, which in simulation form cluster (super)solids.
We also vary the particle mass, i.e., the quantumness of the system, so as to modify the relative stability of ``solid'' and fluid states.
We thus find a whole spectrum of (finite-size) phases with a curved structure, including multilayer and cluster phases, as well as superfluids and single layer supersolids.
In particular, we confirm that supersolids only occur in the presence of clusters.
By considering three system sizes (from 60 to 180 particles), but rescaling the trap radius so that the density stays fixed, we verify that supersolid states only exist in the same range of particle masses.
Our model serves as a platform to test how emergent properties are modified, in spherical geometry, in the crossover from atoms to dressed atoms.
\end{abstract}
\date{\today}
\maketitle

\section{Introduction}

Since the very first observation of Bose-Einstein condensation in 1995~\cite{anderson1995,davis1995}, ultracold atoms have been recognized as a valuable testbed for exploring the emergent behavior of many-body systems~\cite{bloch2008a,bloch2012}.
Continuous progress in trapping and cooling techniques now makes it possible to study particles with highly tunable interactions in a wide variety of confining geometries, such as harmonic wells, optical lattices~\cite{greiner2002}, box potentials~\cite{gaunt2013bose}, and lower-dimensional traps~\cite{gorlitz2001}.
At the same time, a lot of effort has been put in the realization of condensates in curved geometries.
This includes bosons in ring traps~\cite{sengupta2005,morizot2006ring}, in toroidal traps~\cite{ryu2007}, and in bubble traps~\cite{zobay2001,zobay2004}.
In the latter case, which is the one of interest here, the trapping potential is obtained by dressing bosonic (e.g., $^{87}$Rb) atoms in the ground state with a radio-frequency field, in the presence of a quadrupolar magnetic field aligned along $z$ (see more, for example, in \cite{garraway2016,dubessy2025quantum}).
Recently, an all-optical method has been proposed, in which the radio-frequency field is replaced by an optical field~\cite{veyron2026all}.

The main limitation of bubble-trap experiments is the impact of gravity, which pushes bosons to the bottom of the trap, preventing an even distribution of them in the shell.
The eventual realization of shell-shaped condensates in microgravity~\cite{lundblad2019,aveline2020,lundblad2023,carollo2022} has fueled renewed interest in curved confinement, leading to advances in the theory of weakly-interacting bosons on a sphere~\cite{tononi2019}, as well as to a better comprehension of correlation effects like cluster supersolidity in soft-core bosons~\cite{prestipino2019a,ciardi2024,ciardi2025,cinti2026}.
Generally speaking, a system of ultracold bosons in a spherically-symmetric trap offers the exclusive advantage of investigating the interplay of geometry, dimensionality, and many-body physics in one place, opening the way to discoveries and surprises.

Bosons in curved geometry also appear in another context, namely in so-called snowballs of Helium atoms, forming around ionic impurities in bulk superfluid Helium~\cite{Buzzacchi2001,Galli2011,Tramonto2015}.
These assemblies are more correlated than the bulk, as they consist of spherical layers made up of individual atoms, while the outermost layers exhibit transitional supersolidity~\cite{Matos2025}.
The resulting structures show strong similarities to supersolids of soft-core bosons in a broad spherical shell~\cite{cinti2026}, where particles are gathered in clusters, in turn arranged in concentric shells, often exhibiting the symmetries of Platonic polyhedra.
Clearly, Helium atoms are impeded from forming clusters by a harshly repulsive core;
moreover, a supersolid phase has never been found for bulk Helium~\cite{kim2012}.
Therefore, in spite of the existing analogies, there are also marked differences between the two kind of systems, due to distinct interaction regimes.
It would be interesting to see how these differences arise when gradually switching from one system to the other.

To clarify this matter, we perform path integral Monte Carlo simulations of shell-trapped bosons interacting through a potential tuned from a hard-core, Helium-like regime to the soft-core regime typical of Rydberg-dressed atoms.
Using this strategy, we are able to explore collective properties as the nature of the system varies between these two limits.
We note, as a caveat, that although a bounded potential could in principle be realized with Rydberg-dressed atoms \cite{henkel2010,balewski2014,browaeys2020}, in practice making a condensate of excited Rydberg atoms runs into obstacles (due to blackbody radiation induced decay), and putting them in a bubble trap would be even more difficult --- so far, supersolids have only been described in systems of dipolar atoms~\cite{bottcher2019,chomaz2019,tanzi2019,recati2023}.

This paper is organized as follows.
We first describe the model and the simulation method used for its study (Section II).
In Section III we comment our results, explaining how and why the arrangement of atoms changes with the mass and stiffness of the core.
The final Section IV is devoted to conclusions.

\section{Model and Methods}

\begin{figure}[t]
\includegraphics[width=\linewidth]{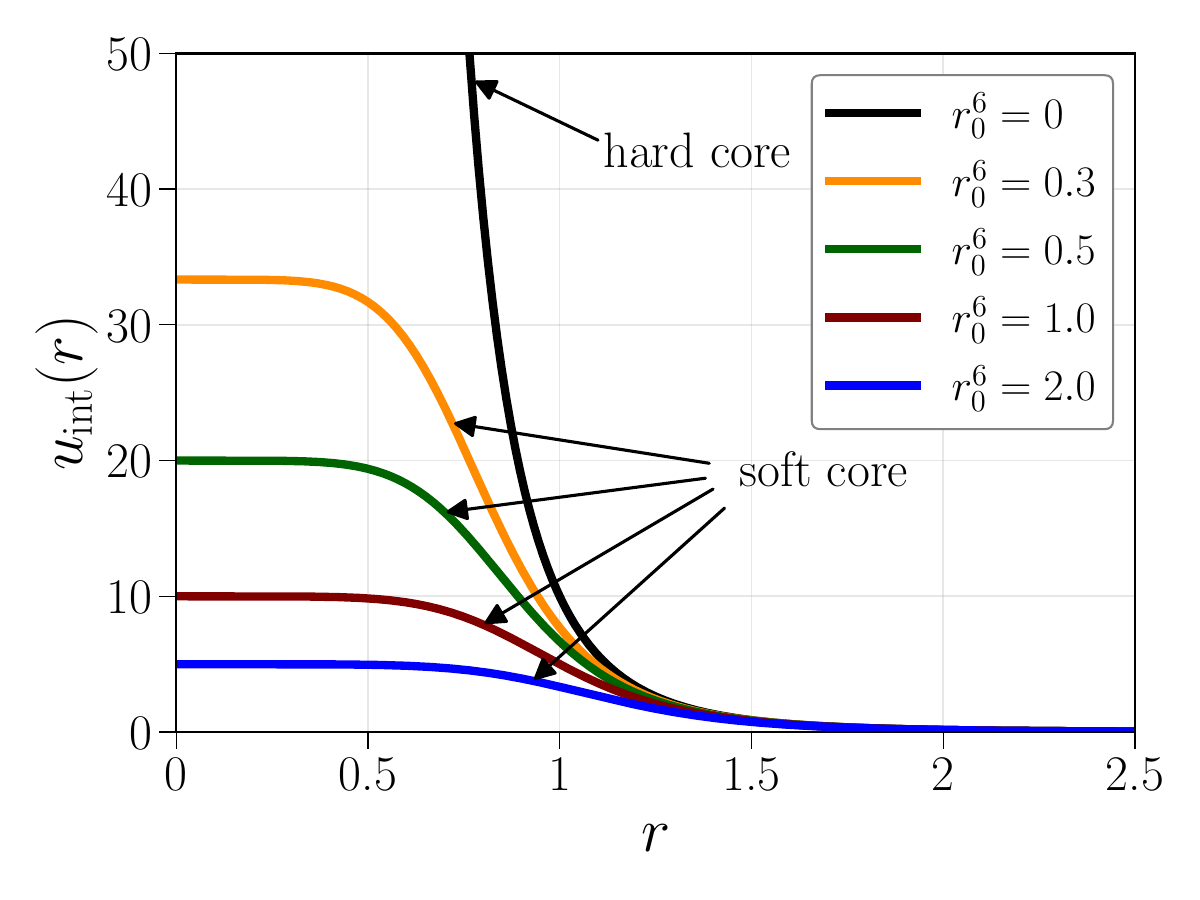}
\caption{Interaction potential (\ref{eq3}) for various $r_0$ values (in the legend).
For $r_0\ne 0$, the inflection point of $u_{\rm int}(r)$ falls at $r=(5/7)^{1/6}r_0=0.9454\ldots r_0$.}
\label{fig1}
\end{figure}

In this paper we study a system of bosons described by the following Hamiltonian:
\begin{equation}
H=\sum_{i=1}^N [-\lambda\nabla_i^2+u_{\textrm{ext}}(|\textbf{r}_i|)]+\sum_{i<j}u_{\textrm{int}}(|\textbf{r}_i-\textbf{r}_j|)\,,
\label{eq1}
\end{equation}
where $\textbf{r}_i$ is the position of the $i$th particle relative to the center of the trap ($i=1,\ldots,N$) and $\lambda=\hbar^2/2m$ measures the strength of quantum fluctuations.
The trapping potential is 
\begin{equation}
u_{\textrm{ext}}(r)=\frac{u_0}{\Omega}\sqrt{\frac{(r^2-R^2)^2}{4}+\Omega^2}-u_0\,,
\label{eq2}
\end{equation}
where $R$ is the trap radius in the thin-shell limit, while $u_0$ and $\Omega$ are other qualifying parameters of the trap~\cite{sun2018}.
In the following, all quantities are given in arbitrary energy and length units.
As boson-boson potential we choose
\begin{equation}
u_{\textrm{int}}(r)=\frac{C}{r_0^6+r^6}\,,
\label{eq3}
\end{equation}
where $r_0$ rules the stiffness of the particle core (see Fig.~\ref{fig1}).
When $r_0\rightarrow 0$ we recover from (\ref{eq3}) a hard-core-like interaction.
As $r_0^6$ is increased in small steps, $u_{\textrm{int}}(0)$ is progressively reduced and the softness of the interaction is then enhanced.
This allows us to investigate how structural and quantum properties change when moving gradually from the Helium-type regime ($r_0=0$), where the bulk solid is an ordinary crystal, to the Rydberg-dressed regime ($r_0\approx 1$), where in two~\cite{cinti2014a,macri2014b} and (in all probability also in) three dimensions atoms solidify into a cluster crystal.
From now onward, $C$ is taken to be 10.

We start our analysis by simulating $N=60$ bosons in a shell trap of radius $R=1$, but we shall later consider larger systems with $120$ or $180$ bosons.
To make a meaningful comparison between the three systems, we require that the average separation between neighboring particles be roughly the same for all sizes.
If particles are distributed in a single spherical layer, this amounts to an equal number density $\rho=N/(4\pi R^2)$.
Hence, we set $R=\sqrt{2}$ for $N=120$ and $R=\sqrt{3}$ for $N=180$.
Clearly, in rescaling the radius we must also ensure that the confining field looks identical to particles near $r=R$.
This implies changing the other parameters too (namely, $u_0$ and $\Omega$) in a proper way.
By Taylor-expanding the potential (\ref{eq2}) around its point of minimum, we obtain:
\begin{multline}
u_{\textrm{ext}}(r)\approx u_0 \Biggl[ \frac{R^2(r-R)^2}{2\Omega^2} + \frac{R(r-R)^3}{2\Omega^2} \\
+ \frac{(\Omega^2-R^4)(r-R)^4}{8\Omega^4} + \ldots \Biggr]\,.
\label{eq4}
\end{multline}

\begin{figure}[H]
\includegraphics[width=\linewidth]{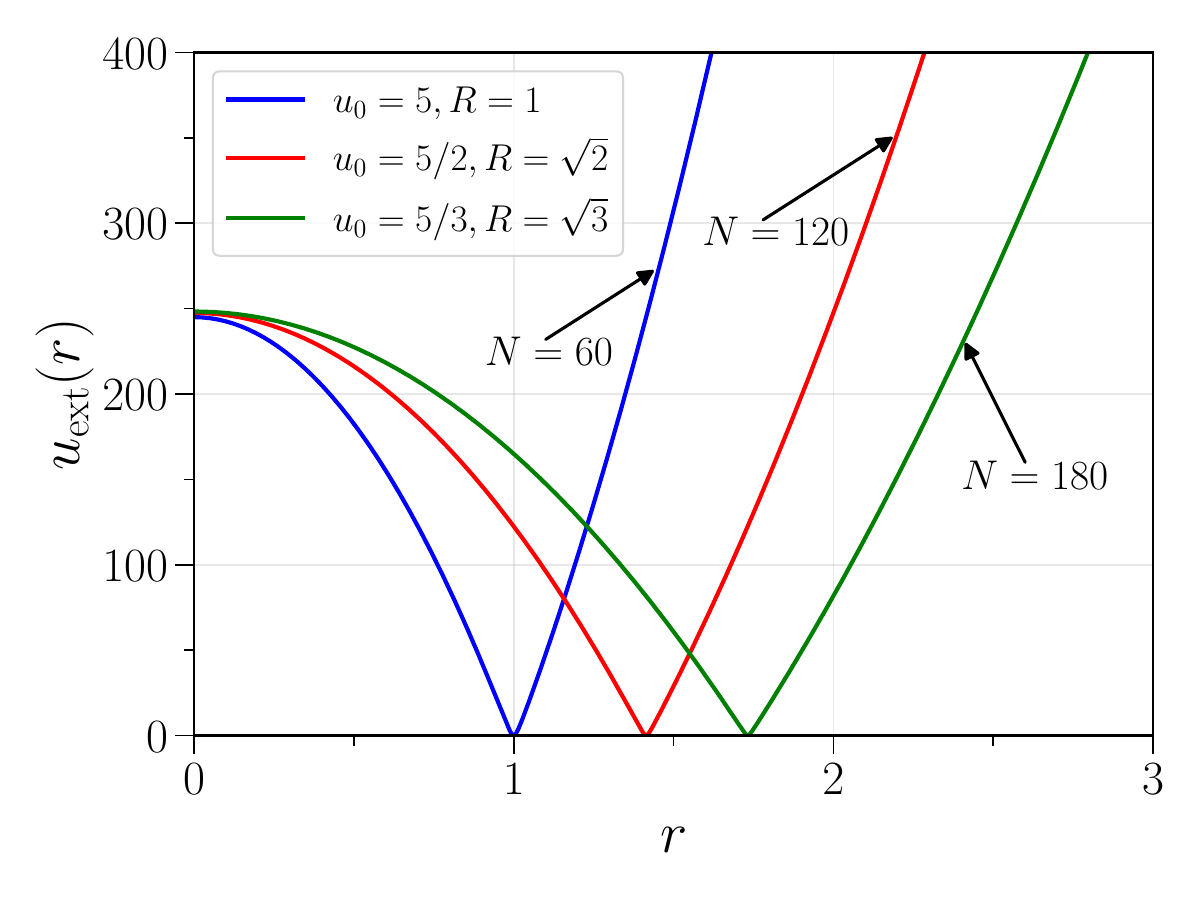}
\caption{Trapping potential for the three system sizes considered in this paper.
Notice the vertical scale.}
\label{fig2}
\end{figure}

We see that rescaling $R$ by a factor of $\alpha$ will not alter the second-order term (for a given departure from the bottom of the well) only provided that $u_0/\Omega^2$ is simultaneously reduced by $\alpha^2$.
Taking a fixed $\Omega$ and rescaling $u_0\rightarrow u_0/\alpha^2$, we see from Eq.~(\ref{eq4}) that the third-order term will then be reduced by $\alpha$ and the fourth-order term (approximately) multiplied by $\alpha^2$.
Therefore, it is impossible to adjust the trap parameters so that, say, both the second- and third-order terms stay unchanged.
In conclusion, we cannot do better than keeping the coefficient of $(r-R)^2$ identical for all sizes.
With this agreement, the trapping potentials for the three sizes are those displayed in Fig.~\ref{fig2}.

For our simulations we employ the path integral Monte Carlo (PIMC) method~\cite{ceperley1995}, using the worm algorithm~\cite{boninsegni2006a,boninsegni2006b} for an efficient sampling of particle permutations.
Our simulations are canonical, i.e., they are performed at a fixed temperature and particle number. Here, we choose a low temperature ($T=1$).
The PIMC method rests upon the isomorphism between quantum particles and classical imaginary-time paths (worldlines).
For boson systems at equilibrium, the worldlines of individual particles are allowed to join together, thus forming closed chains of variable length $L$ (the number of constituent worldlines).
In this regard, we compute the probability $P(L)$ that a particle chosen at random belongs to a chain/cycle of length
$L$~\cite{jain2011a,ciardi2024,ciardi2025}.
While a slow decay of $P(L)$ would be suggestive of superfluid behavior, a sharp drop in the profile points to the existence of clusters.

The superfluid fraction $f_s$ is evaluated by sampling the ``area estimator''~\cite{ceperley1995,sindzingre1989}, which relates the variance of the (projected) area enclosed by the chains to the reduction of the moment of inertia with respect to its classical value.
In particular, we probe the superfluid fraction $f_s^{(k)}$ along three mutually orthogonal axes ($k=x,y$, and $z$).
In doing so, the superfluid fraction reads:
\begin{equation}
f_s^{(k)}=\frac{2\beta}{\lambda}\frac{\langle A_k^2\rangle-\langle A_k\rangle^2}{I_{\rm cl}^{(k)}}\,.
\label{eq5}
\end{equation}
In this expression, $I_{\rm cl}^{(k)}$ is the classical moment of inertia relative to the $k$th direction and $A_k$ is the total area enclosed by the chains, projected onto the plane perpendicular to the $k$ axis.
Given the isotropic character of the interaction (\ref{eq3}), the superfluid fractions (\ref{eq5}) are equal.
Therefore, we report the single value $f_s=(f_s^{(x)}+f_s^{(y)}+f_s^{(z)})/3$.
While in a bulk system a non-zero $f_s$ indicates that the state is globally phase-coherent, hence superfluid, in our (inhomogeneous, finite-size) sample a small $f_s$, say below 0.1, would be more correctly associated with a normal/insulating state, either solid or cluster-solid.

We also compute the radial distribution function (RDF) $g(r)$, where $r$ is the Euclidean distance in three dimensions.
In a PIMC simulation, $g(r)$ is sampled as a histogram.
The interparticle distance is divided in bins, and a reference particle is chosen;
at each sampling step, the distance between the reference particle and every other particle is computed, and the histogram at the corresponding bin is updated.
Sampling is done at each time slice and further averaged over time slices.
Generally speaking, clusterization causes $g(0)$ to be non-zero (the more populated the clusters, the higher the contact value); moreover, a sharper solidity comes along with more pronounced RDF peaks beyond the first one.

A quantity which conveys information complementary to that contained in the RDF is the spherical structure factor (SSF)~\cite{bovzivc2021global,lovsdorfer2019spherical}.
In Appendix A we report its definition, as well as a proof-of-principle calculation providing insight into the relationship of the SSF to the symmetries inherent in the cluster structure.
Concrete examples of this link are given in the next section.

\section{Results}

\begin{figure*}[t]
    \centering
    \begin{minipage}[t]{0.2\linewidth} 
        \centering
        \textbf{$r_0^6=0$}\par\medskip
        \includegraphics[width=1.0\linewidth]{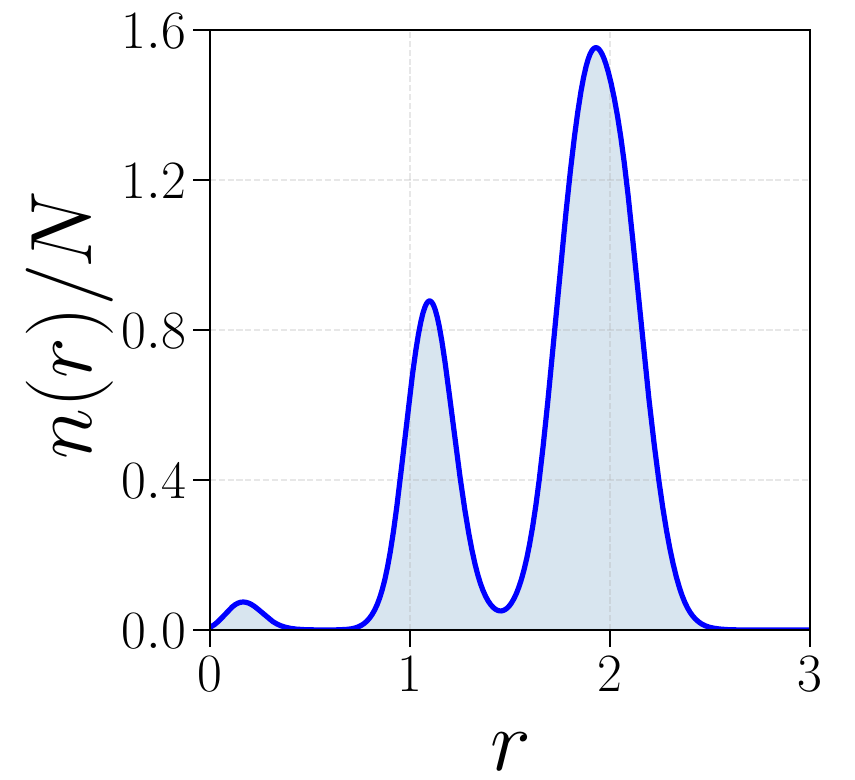}
    \end{minipage}\hfill
    \begin{minipage}[t]{0.2\linewidth} 
        \centering
        \textbf{$r_0^6=0.2$}\par\medskip
        \includegraphics[width=0.88\linewidth]{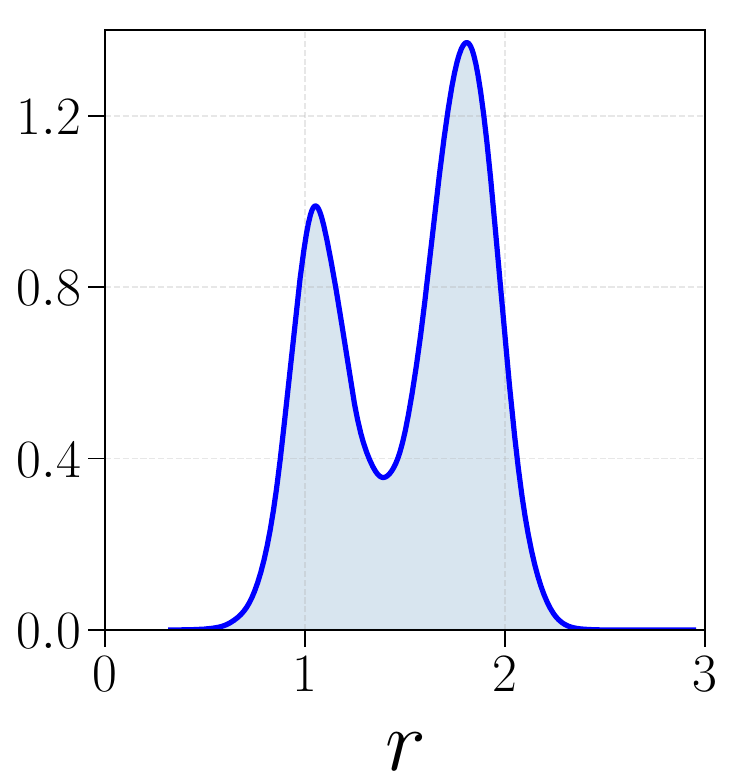}
    \end{minipage}\hfill
    \begin{minipage}[t]{0.2\linewidth} 
        \centering
        \textbf{$r_0^6=0.3$}\par\medskip
        \includegraphics[width=0.88\linewidth]{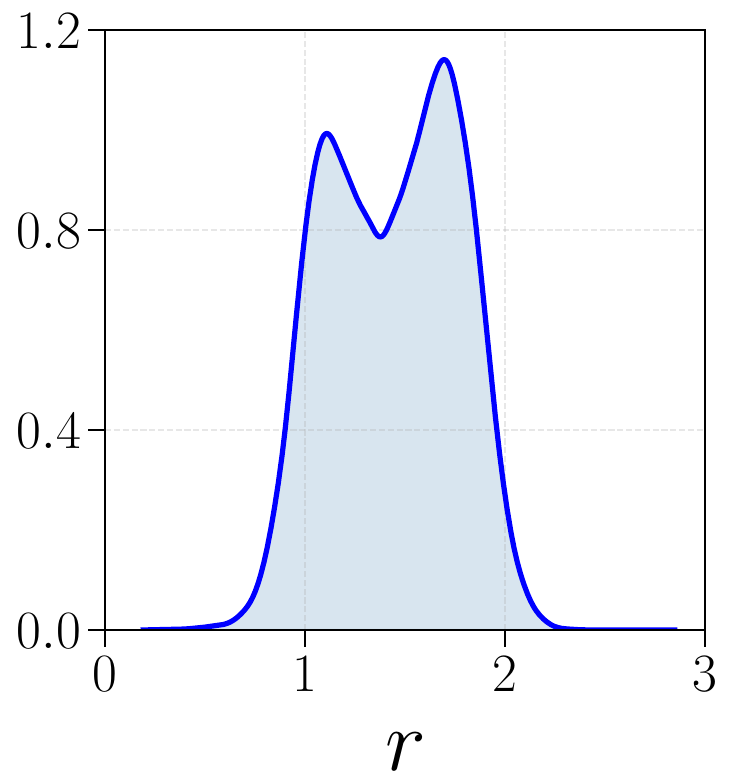}
    \end{minipage}\hfill
        \begin{minipage}[t]{0.2\linewidth}
        \centering
        \textbf{$r_0^6=0.5$}\par\medskip
        \includegraphics[width=0.88\linewidth]{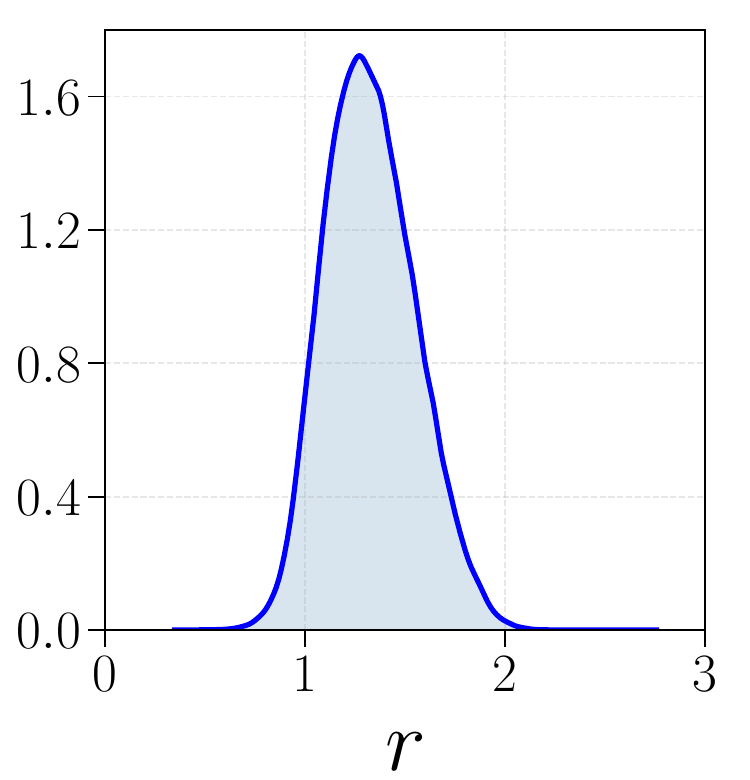}
    \end{minipage}\hfill
    \begin{minipage}[t]{0.2\linewidth} 
        \centering
        \textbf{$r_0^6=3$}\par\medskip
        \includegraphics[width=0.88\linewidth]{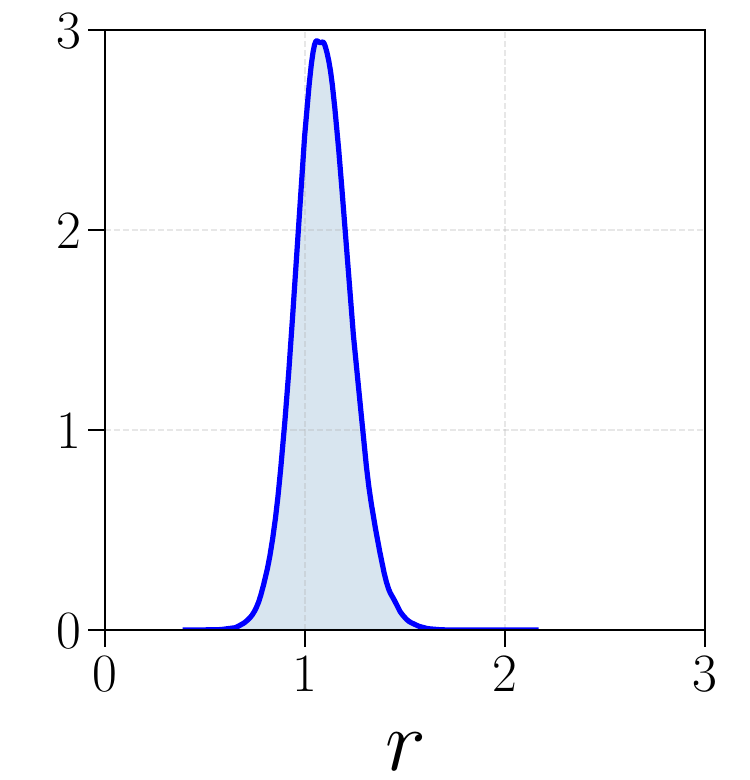}
    \end{minipage}
    \caption{$N=60$ bosons: Integrated radial density $n(r)$, plotted for $\lambda=0.3$ and a few values of the softness parameter $r_0^6$. }
    \label{fig3}
\end{figure*}

We first consider a system of $60$ bosons in a trapping potential with parameters $R=1,u_0=5$, and $\Omega=0.01$ (the blue curve in Fig.~\ref{fig2}).
For each value of $\lambda$ and $r_0$, particles are initially placed at random on the spherical surface of radius $R$ and then equilibrated until all relevant properties have settled.
Statistical averages are typically computed over a span of $5\times 10^8$ elementary PIMC moves.
In Fig.~\ref{fig3} we plot, for $\lambda=0.3$ and various $r_0^6$, the integrated radial density $n(r)$ divided by $N$~\cite{ciardi2025}.
We see that, in the hard-core limit ($r_0=0$), particles are not located near $r=R$.
Rather, they are distributed in two concentric spherical layers, respectively populated with $16$ and $43$ particles, surrounding one central particle --- a structure reminding of helium snowballs.
As the interaction becomes increasingly softer, the two layers eventually merge into one, which then progressively shifts toward $r=R$. 
This behavior is somewhat similar to that reported for a square-well interaction in Ref.~\onlinecite{ciardi2025}, where  a gradual increase of $N$ drives the reverse transformation (from particle clusters in one layer to clusters distributed in two layers).
In the present case, it is the $r^{-6}$ repulsion that induces a fraction of the particles to climb up the external potential and form a second spherical layer at larger distance from the trap center.
As $r_0^6$ overcomes $\approx 0.3$, the balance between the internal and the external potential energy changes, and the optimal choice in energetic terms becomes forming a  single layer.

\begin{figure*}[t]
    \centering
    \begin{minipage}[t]{0.5\linewidth} 
        \centering
        \textbf{(a)\hspace{10pt}$\lambda=0.2,r_0^6=0$}\par\medskip
        \includegraphics[width=1.0\linewidth]{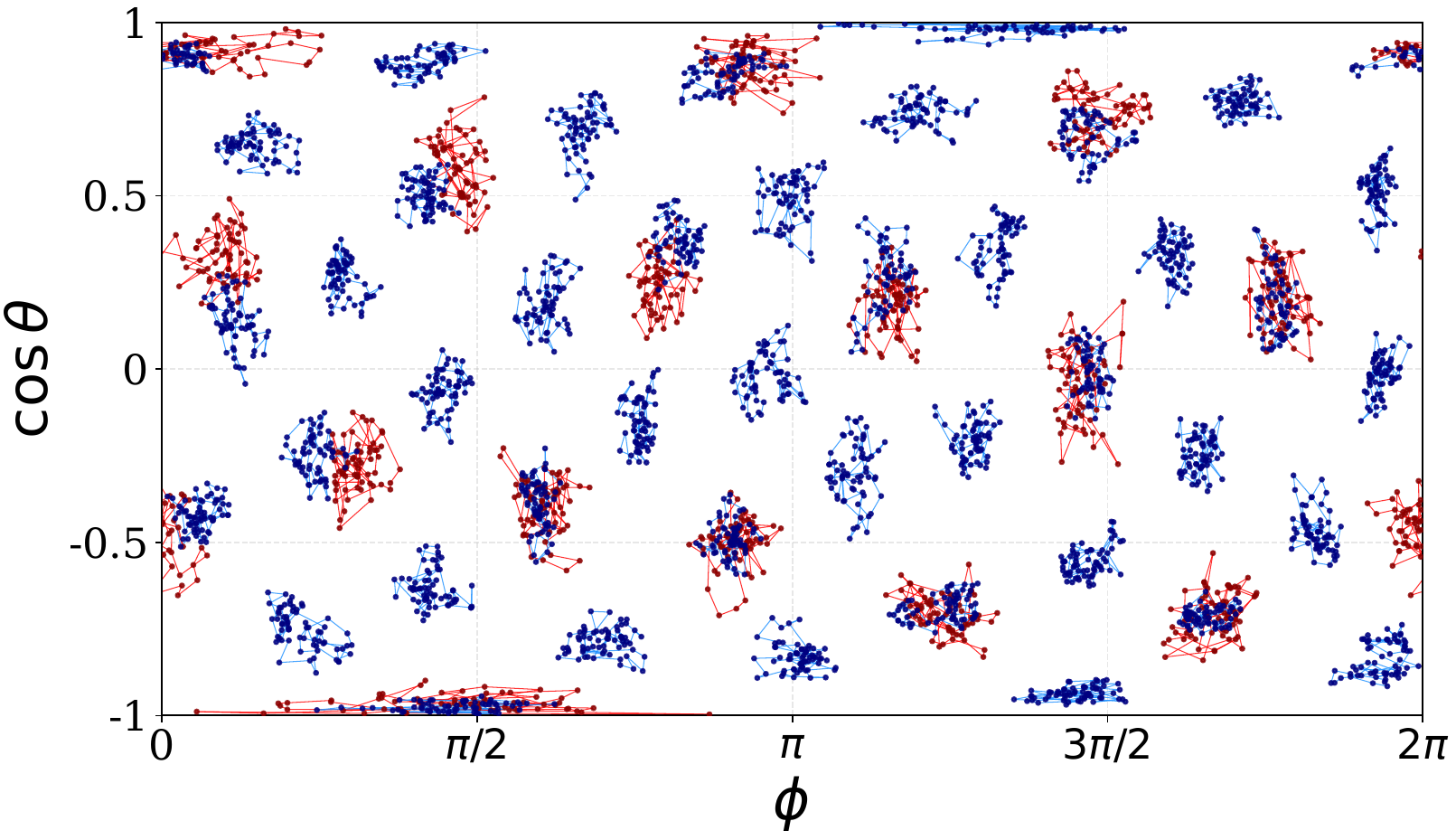}
    \end{minipage}\hfill
    \begin{minipage}[t]{0.5\linewidth} 
        \centering
        \textbf{(b)\hspace{10pt}$\lambda=0.15,r_0^6=0.6$}\par\medskip
        \includegraphics[width=0.955\linewidth]{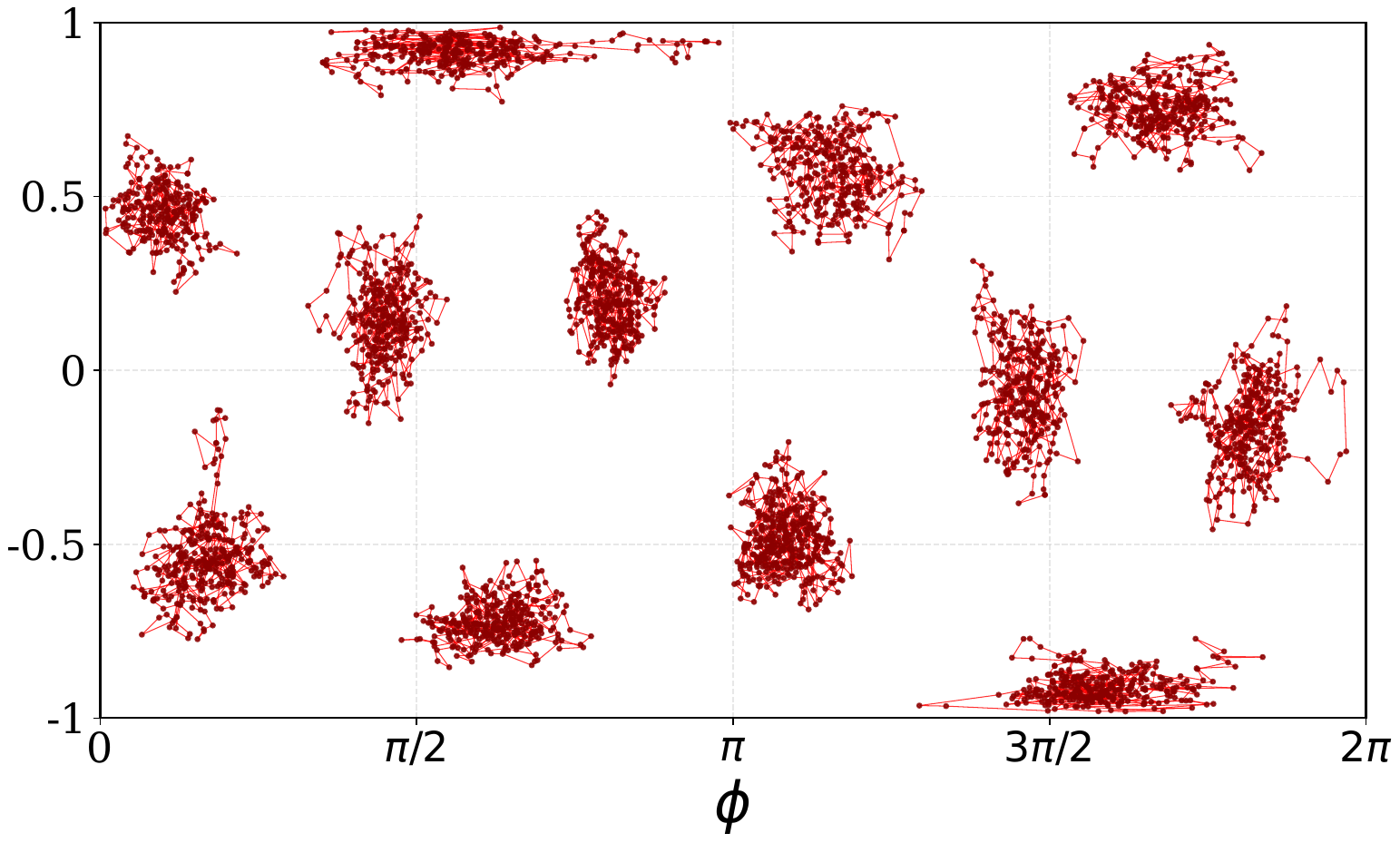}
    \end{minipage}\\
    \vspace{0.5cm}
    \begin{minipage}[t]{0.5\linewidth} 
        \centering
        \textbf{(c)\hspace{10pt}$\lambda=0.3,r_0^6=1.8$}\par\medskip
        \includegraphics[width=1.0\linewidth]{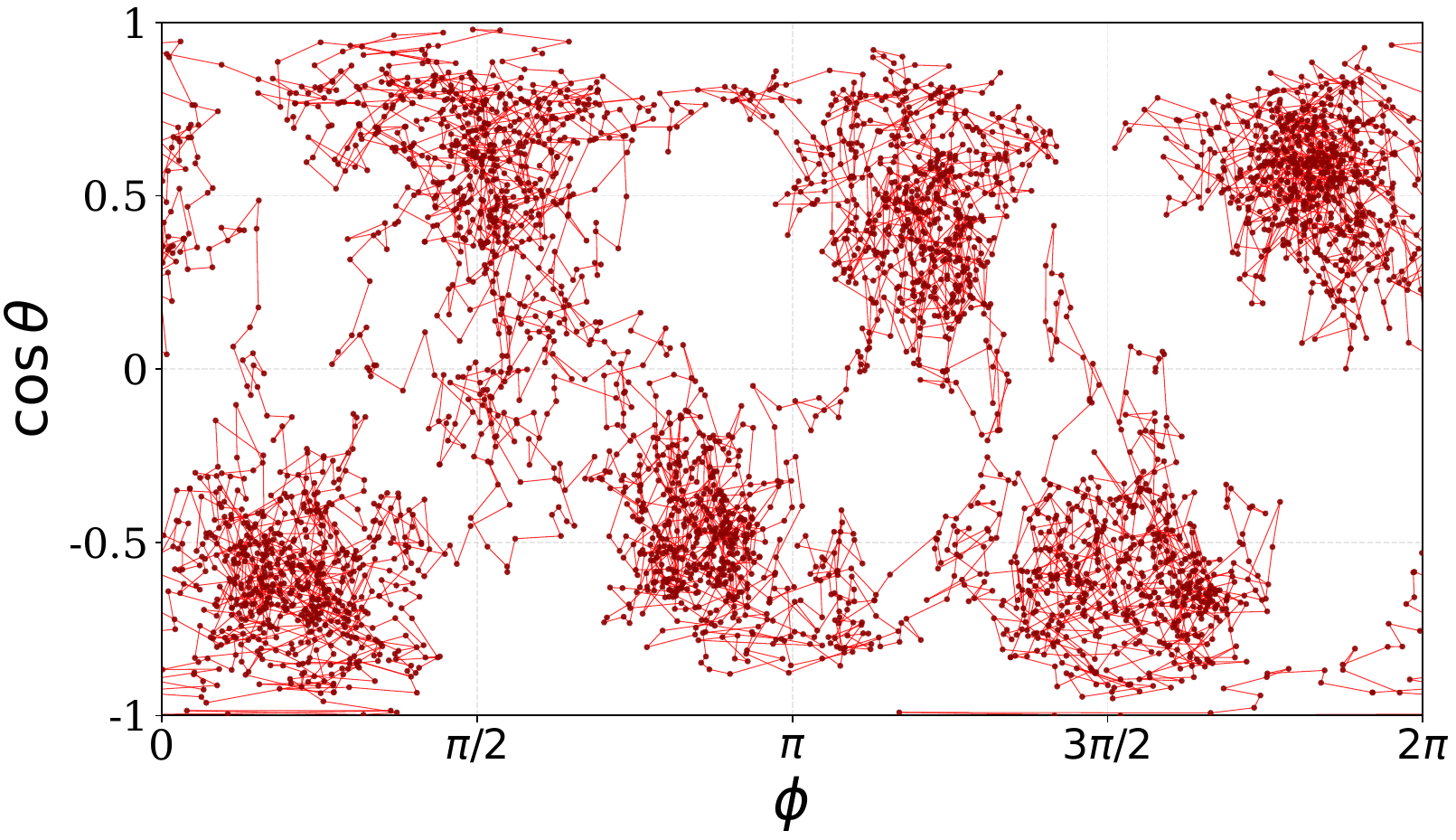}
    \end{minipage}\hfill
        \begin{minipage}[t]{0.5\linewidth}
        \centering
        \textbf{(d)\hspace{10pt}$\lambda=0.5,r_0^6=2$}\par\medskip
        \includegraphics[width=0.955\linewidth]{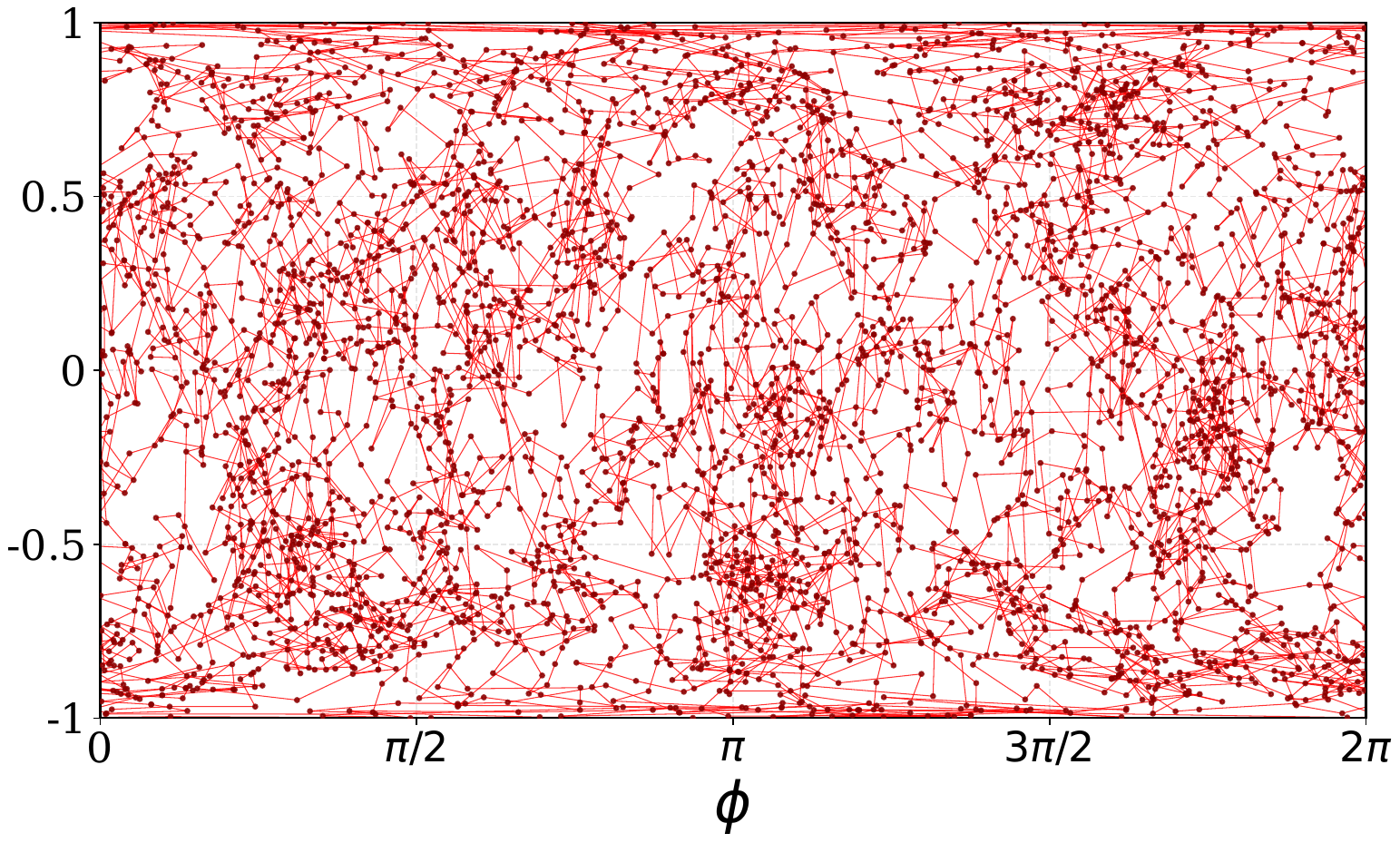}
    \end{minipage}
    \caption{$N=60$ bosons: Four typical bead configurations (projection maps). (a) Solid bilayer (red: particles in the inner layer; blue: particles in the outer layer); (b) Icosahedral cluster solid; (c) Octahedral cluster supersolid; and (d) Superfluid.}
    \label{fig4}
\end{figure*}

\begin{figure*}[t]
\includegraphics[width=\linewidth]{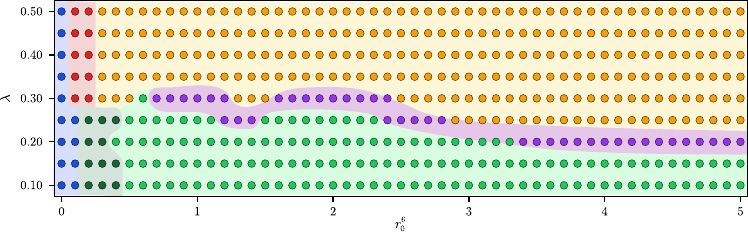}
\includegraphics[width=0.45\linewidth]{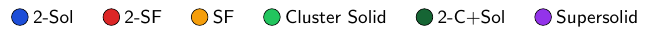}
\caption{$N=60$ bosons: Phase diagram at $T=1$. The color palette of the circles is as follows: blue $\rightarrow$ solid bilayer (2-Sol); red $\rightarrow$ superfluid bilayer (2-SF); yellow $\rightarrow$ superfluid (SF); light green $\rightarrow$ cluster solid; dark green $\rightarrow$ bilayer with clusters in the first layer and isolated particles in the second layer (2-C$+$Sol); purple $\rightarrow$ supersolid. Background colors and phase boundaries are a guide for the eye.}
\label{fig5}
\end{figure*}

To have a clearer view of what is happening in the system we may look at two-dimensional maps of the bead configuration (akin to Lambert cylindrical projections of the sphere surface).
The positions of particle beads are plotted as $(\phi,\cos\theta)$ pairs, where $\theta$ is the polar angle and $\phi$ is the azimuthal angle.
A few examples are shown in Fig.~\ref{fig4}, where, depending on $r_0$ and $\lambda$, particles either stay well separate, form more or less compact clusters, or spread uniformly throughout the layer(s).
Clearly, visual inspection of individual configurations cannot be the only criterion, nor even the main one, to identify the nature of the system.
In fact, it is primarily from the value of the superfluid fraction that we get an idea of the character of the system for given $r_0$ and $\lambda$.
In particular, a system with $f_s\lesssim 0.1$ is classified as ``normal'' (i.e., non-superfluid), whereas one with $f_s\gtrsim 0.65$ is ``superfluid''.
States with intermediate values of $f_s$ are dubbed ``supersolid'' if structural indicators (RDF and SSF), as well as the cycle-length distribution and visual inspection, all point to the existence of clusters.
The reason for assuming a non-zero lower bound for $f_s$ of a supersolid state is the superfluidity of the individual clusters. Since the number of clusters in a finite solid sample is relatively small, the internal superfluidity of a single cluster will still give a small non-zero contribution to the overall superfluidity.
As arbitrary as the above classification rules may appear, they allow us to recognize physically distinct situations, which are then arranged in a ``phase diagram'' (see Fig.~\ref{fig5}).
Clearly, these ``phases'' are finite-size characteristics which, by their very nature, only exist in the presence of the trap.
Moreover, the sequence of phases would be different at a higher temperature, where the free-energy balance between solid and (super)fluid generally shifts towards the latter.

Looking more closely at Fig.~\ref{fig5}, the system properties vary considerably with $r_0^6$ and $\lambda$, especially at the lowest $r_0$ values and for $\lambda\approx 0.3$.
While a harsh core repulsion pushes particles apart and causes the onset of a multilayer structure, a soft enough repulsion leads to cluster formation, at least provided that quantum fluctuations are not too strong.
For values of $r_0^6$ between 0.2 and 0.4, clusters are relegated to the inner layer, while isolated particles still populate the outer layer.

\begin{figure}[t]
\centering
\includegraphics[width=1.0\linewidth]{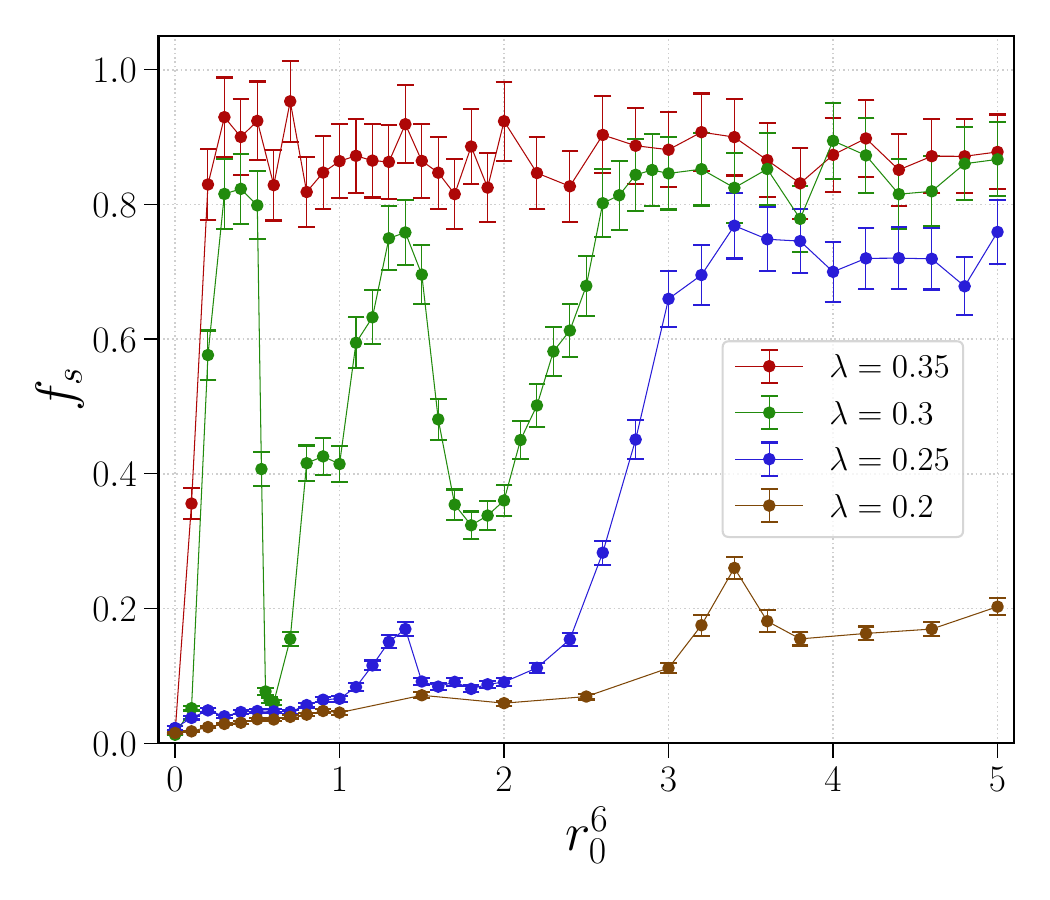}
\caption{$N=60$ bosons: Superfluid fraction for a number of $\lambda$ values (in the legend), plotted as a function of $r_0^6$.}
\label{fig6}
\end{figure}

At the boundary between superfluid and cluster states stands out a thin island of supersolidity, which bends downward to lower $\lambda$ as $r_0^6$ increases beyond $\approx 3$.
As can be appreciated from the non-monotonous evolution of $f_s$ for $\lambda=0.3$ as a function of $r_0^6$ (Fig.~\ref{fig6}), for this critical value of $\lambda$ the system walks a fine line between order and disorder, consistent with the twofold nature of the supersolid, which mix together solid features (a spatially-modulated density) and superfluid features (an overall phase coherence).

In general, the number of clusters is weakly dependent on $\lambda$, whereas it varies strongly with $r_0^6$, becoming smaller with increasing $r_0^6$.
For example, the number of clusters is 12, 9, and 6 for $r_0^6=0.6,1$, and 1.8, respectively.
In particular, 12 clusters typically occur at the vertices of an icosahedron, while 6 clusters lie at the vertices of an octahedron.
We have also observed a tetrahedral arrangement of clusters, for $r_0^6\approx5$ and $\lambda\leq 0.2$.

\begin{figure}[t]
\includegraphics[width=\linewidth]{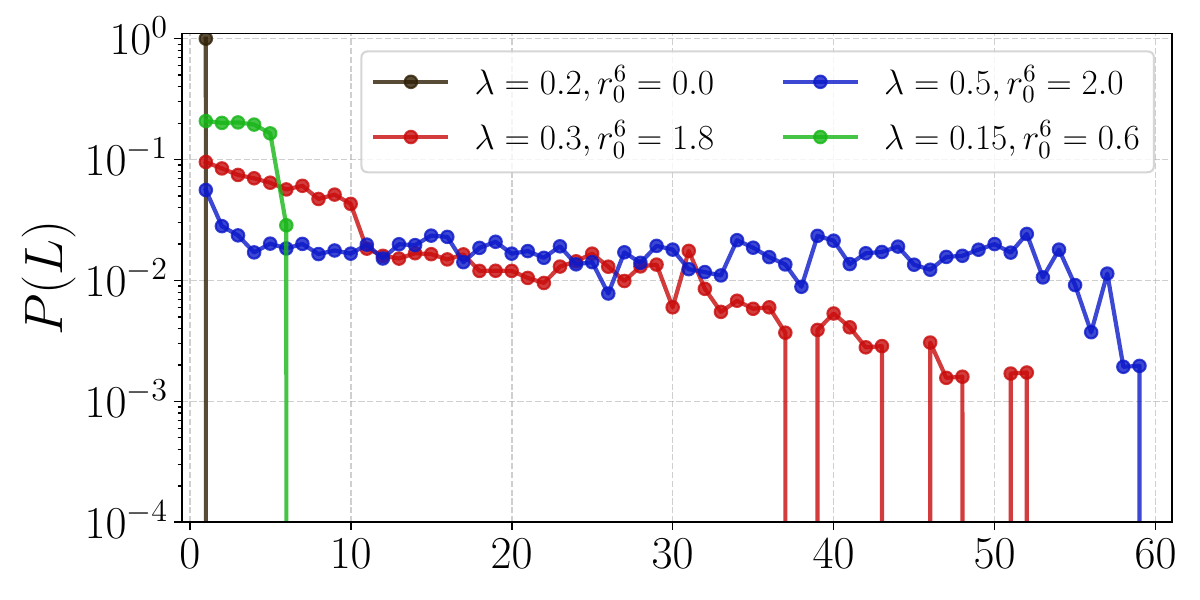}\\
\includegraphics[width=\linewidth]{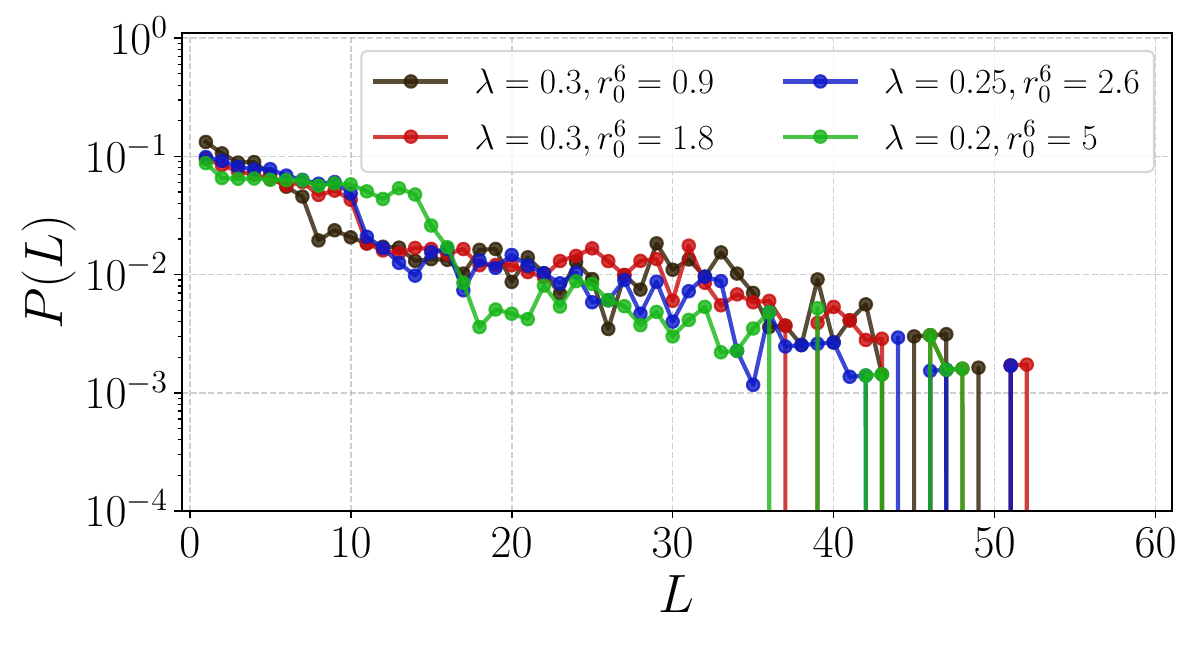}
\caption{$N=60$ bosons.
Top: Statistics of cycle lengths for a few selected states (in the legend).
Bottom: $P(L)$ for various supersolid states (in the legend), with 9 (brown), 6 (red and blue), and 4 clusters (green), respectively. The red curve also appears in the top panel.}
\label{fig7}
\end{figure}

The diversity of boson phases with varying $r_0$ and $\lambda$ is also evidenced in the shape of the distribution $P(L)$ of cycle lengths.
In Fig.~\ref{fig7}a we report $P(L)$ for the same cases illustrated in Fig.~\ref{fig4}.
As anticipated in Section II, a rapidly decaying $P(L)$ is suggestive of a normal, non-superfluid phase.
If $P(L)$ vanishes at a $L$ value (say, $L_0$) larger than 1, then $L_0$ represents the average population of clusters (since up to $L_0$ worldlines can join together).
On the contrary, a non-decaying $P(L)$ is typical of a superfluid state.
Finally, the system is supersolid when $P(L)$ is long-tailed (indicating superfluidity) and we see a drop/step down in its profile (indicating the presence of clusters).
Figure \ref{fig7}a shows that, in a supersolid state, the large-$L$ decay of $P(L)$ is nearly exponential.
This feature is shared by all the supersolids met in our study (see Fig.~\ref{fig7}b).

\begin{figure}[t]
\includegraphics[width=\linewidth]{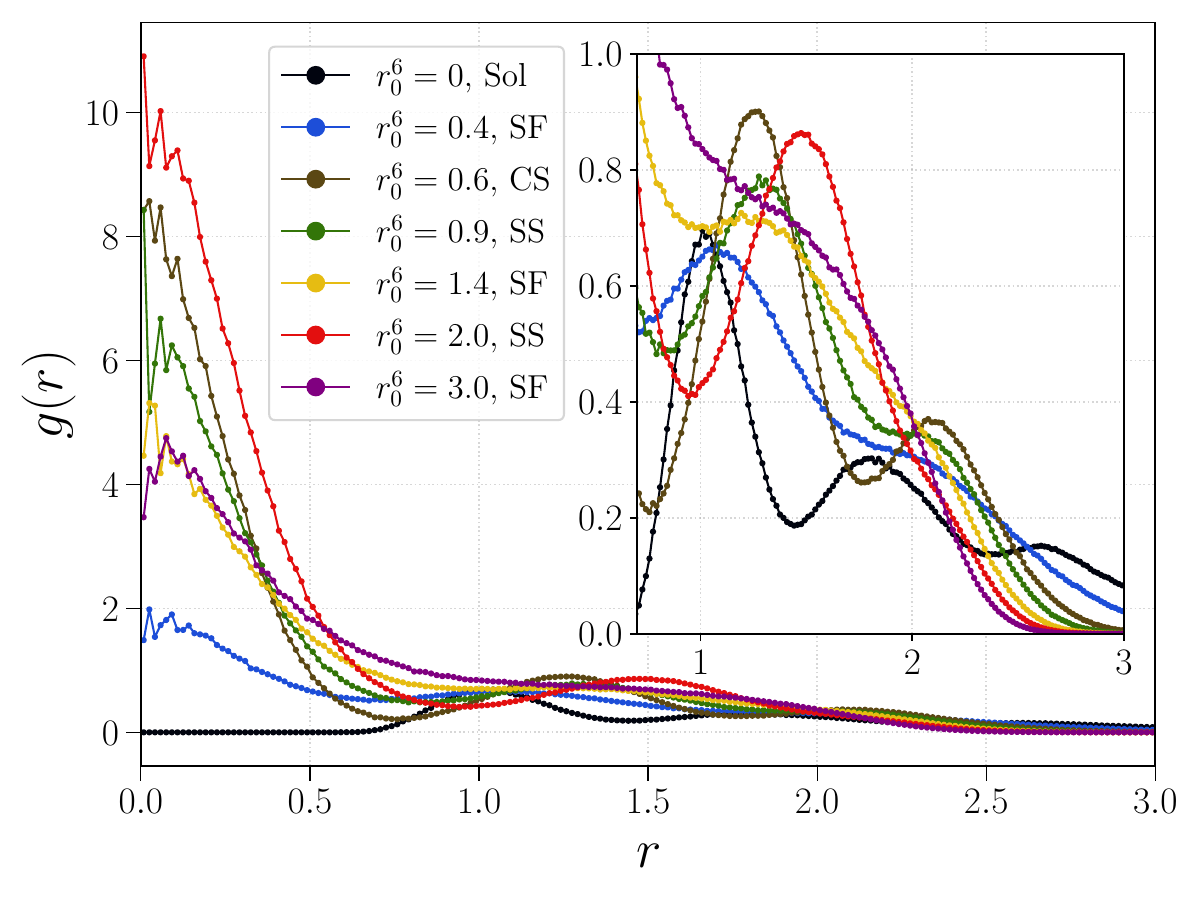}
\caption{$N=60$ bosons: RDF for various states (in the legend).
Inset: magnification of the large-$r$ region. }
\label{fig8}
\end{figure}

In order to distinguish a fluid-like from a solid-like structure we also look at the RDF, plotted in Fig.~\ref{fig8} for several values of $r_0$ at fixed $\lambda=0.3$.
With the exception of the hard-core system ($r_0=0$), all profiles exhibit the non-zero contact value typical of bounded interactions.
The extent of $g(0)$, as well as the existence and prominence of peaks at non-zero $r$, can be used to discriminate between the various phases.
As we see in Fig.~\ref{fig8}, the more populated the clusters the larger $g(0)$.
Moreover, we see in the figure inset that a stronger solidity corresponds to higher RDF peaks.

\begin{figure}[t]
\includegraphics[width=\linewidth]{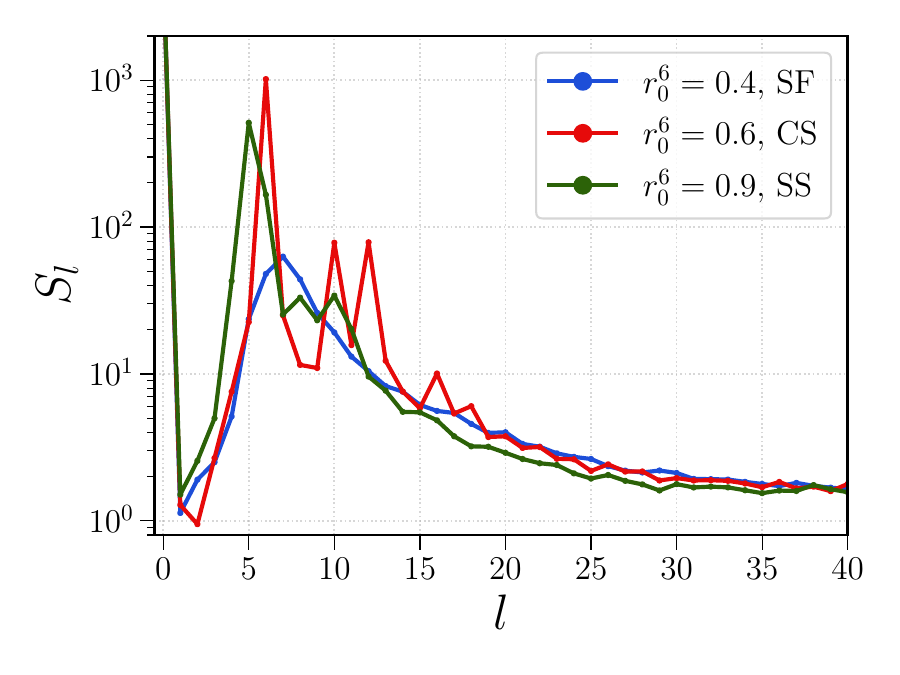}
\caption{$N=60$ bosons: SSF for $\lambda=0.3$ and a few values of $r_0^6$ (in the legend).}
\label{fig9}
\end{figure}

In Fig.~\ref{fig9}, which again refers to $\lambda=0.3$, we show the SSF for three distinct cases, corresponding to a superfluid state ($r_0^6=0.4$), a cluster-solid state ($r_0^6=0.6$, with 12 clusters), and a supersolid state ($r_0^6=0.9$, with 9 clusters).
While the superfluid has the smoother SSF of all, we see peaks in the other two cases, occurring at values of the orbital quantum number $l$ that are sensitive to the geometry of the overall structure.
For example, in the cluster solid (red curve in Fig.~\ref{fig9}) the first SSF peaks are located at $l=0,6,10,12,16,18$, i.e., exactly where they would be expected in a perfect icosahedral structure (see the appendix).
Therefore, we can safely conclude from this picture that the 12 clusters are centered at the vertices of an icosahedron.

\begin{figure*}[t]
    \centering
    \begin{minipage}[t]{0.2\linewidth} 
        \centering
        \textbf{$r_0^6=0$}\par\medskip
        \includegraphics[width=1.0\linewidth]{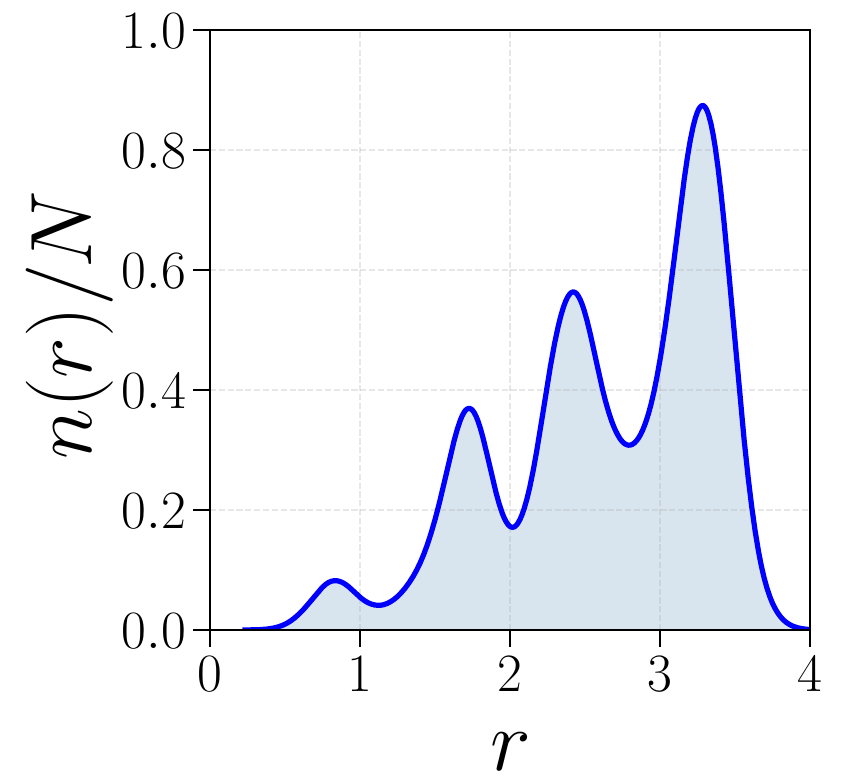}
    \end{minipage}\hfill
    \begin{minipage}[t]{0.2\linewidth} 
        \centering
        \textbf{$r_0^6=0.1$}\par\medskip
        \includegraphics[width=0.88\linewidth]{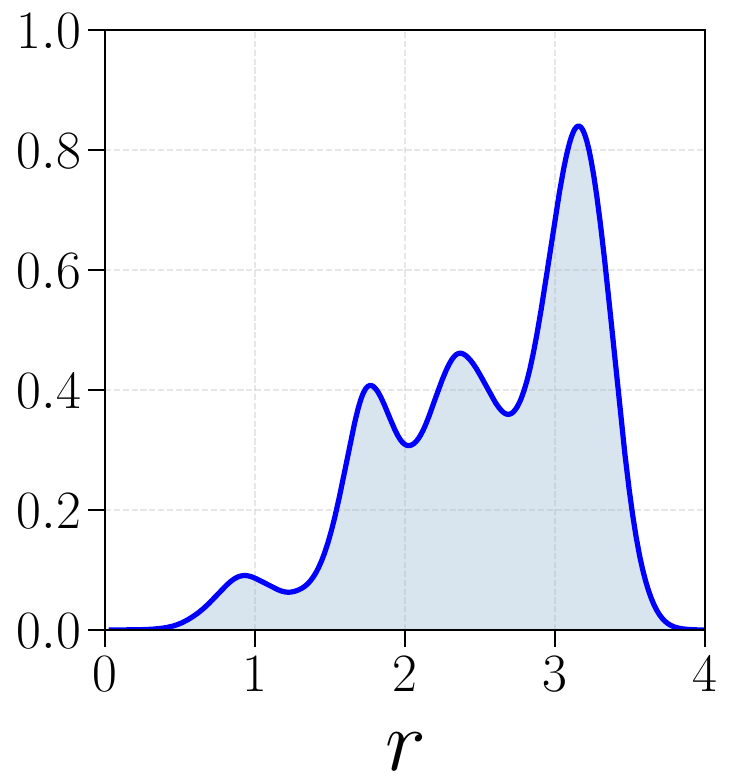}
    \end{minipage}\hfill
    \begin{minipage}[t]{0.2\linewidth} 
        \centering
        \textbf{$r_0^6=0.3$}\par\medskip
        \includegraphics[width=0.88\linewidth]{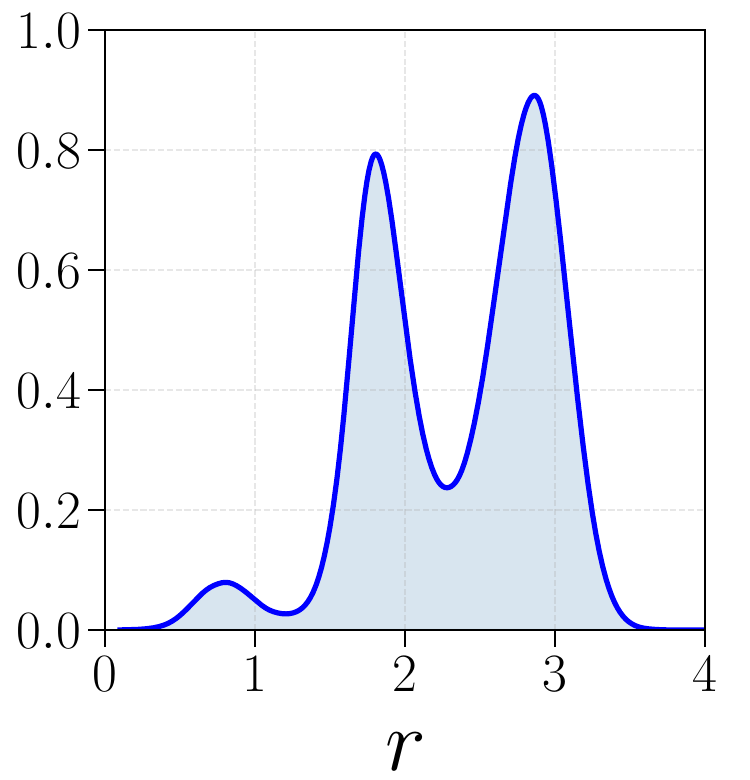}
    \end{minipage}\hfill
        \begin{minipage}[t]{0.2\linewidth}
        \centering
        \textbf{$r_0^6=0.7$}\par\medskip
        \includegraphics[width=0.88\linewidth]{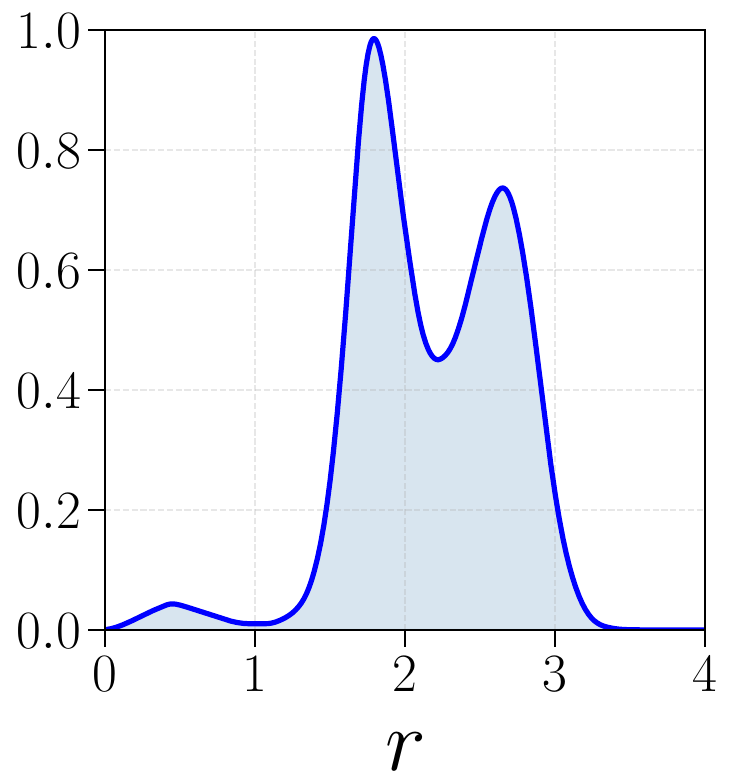}
    \end{minipage}\hfill
    \begin{minipage}[t]{0.2\linewidth} 
        \centering
        \textbf{$r_0^6=3$}\par\medskip
        \includegraphics[width=0.88\linewidth]{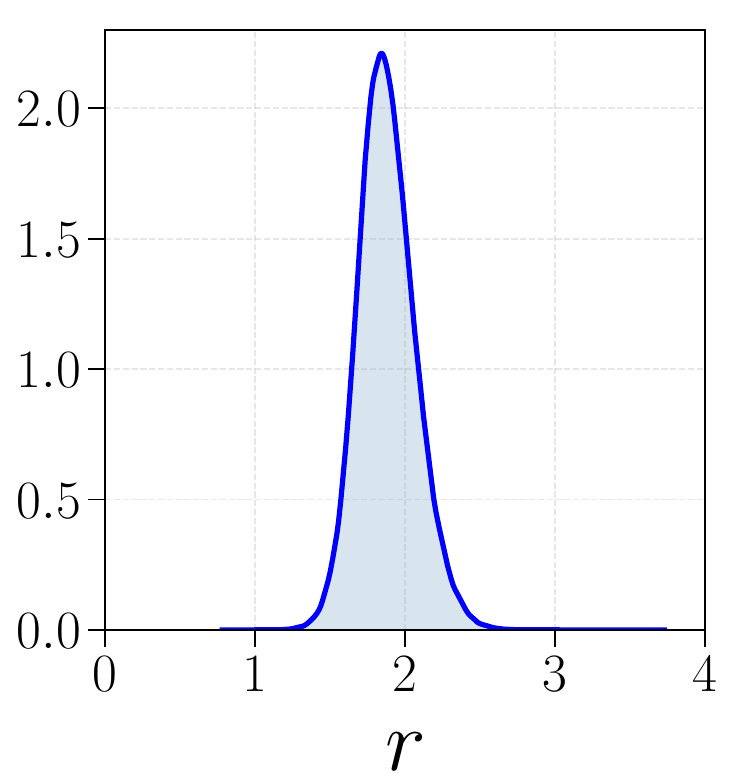}
    \end{minipage}
    \caption{$N=180$ bosons: Integrated radial density $n(r)$, plotted for $\lambda=0.3$ and a few values of the softness parameter $r_0^6$.}
    \label{fig10}
\end{figure*}

Next, we move to bigger samples, i.e., $N=120$ and $N=180$, to see whether the phases found for $N=60$ would survive a rescaling of the system size.
The crucial point to address is how the comparison between different sizes must be carried out in order to be significant.
We adopt the criterion that two one-layer systems are only comparable if they have (roughly) the same specific volume.
However, in this circumstance it is actually impossible to adjust parameters so that the external potential looks the same near the minimum for all sizes.
As noticed in Section II, the best we can do is to make at least the curvature of the external potential equal for all $N$.

\begin{figure*}[t]
\includegraphics[width=\linewidth]{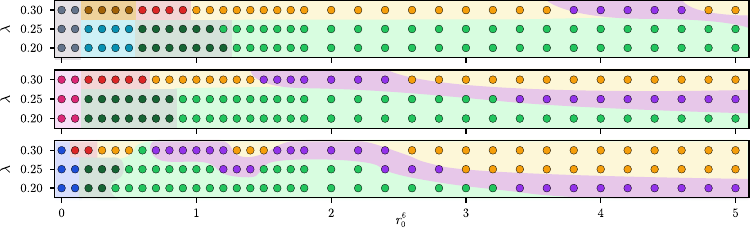}
\includegraphics[width=0.75\linewidth]{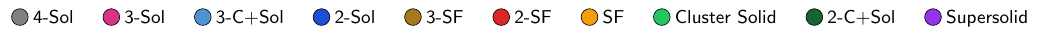}
\caption{We show, in a narrow range of $\lambda$ values, a comparison of phase diagrams for three system sizes (top: $N=180$; middle: $N=120$; bottom: $N=60$).
For the sake of comparison, we have reproduced an excerpt of the phase diagram for $N=60$ reported in Fig.~\ref{fig5}.
We adopt the same color code of Fig.~\ref{fig5}.
Extra colors have the following meaning:
grey $\rightarrow$ solid 4-layer (4-Sol); pink $\rightarrow$ solid trilayer (3-Sol); light blue $\rightarrow$ trilayer with isolated particles in the first and third layers, and clusters in the second layer (3-C$+$Sol); brown $\rightarrow$ superfluid trilayer (3-SF). Background colors and phase boundaries are a guide for the eye.}
\label{fig11}
\end{figure*}

\begin{figure}[H]
\centering
\includegraphics[width=1.0\linewidth]{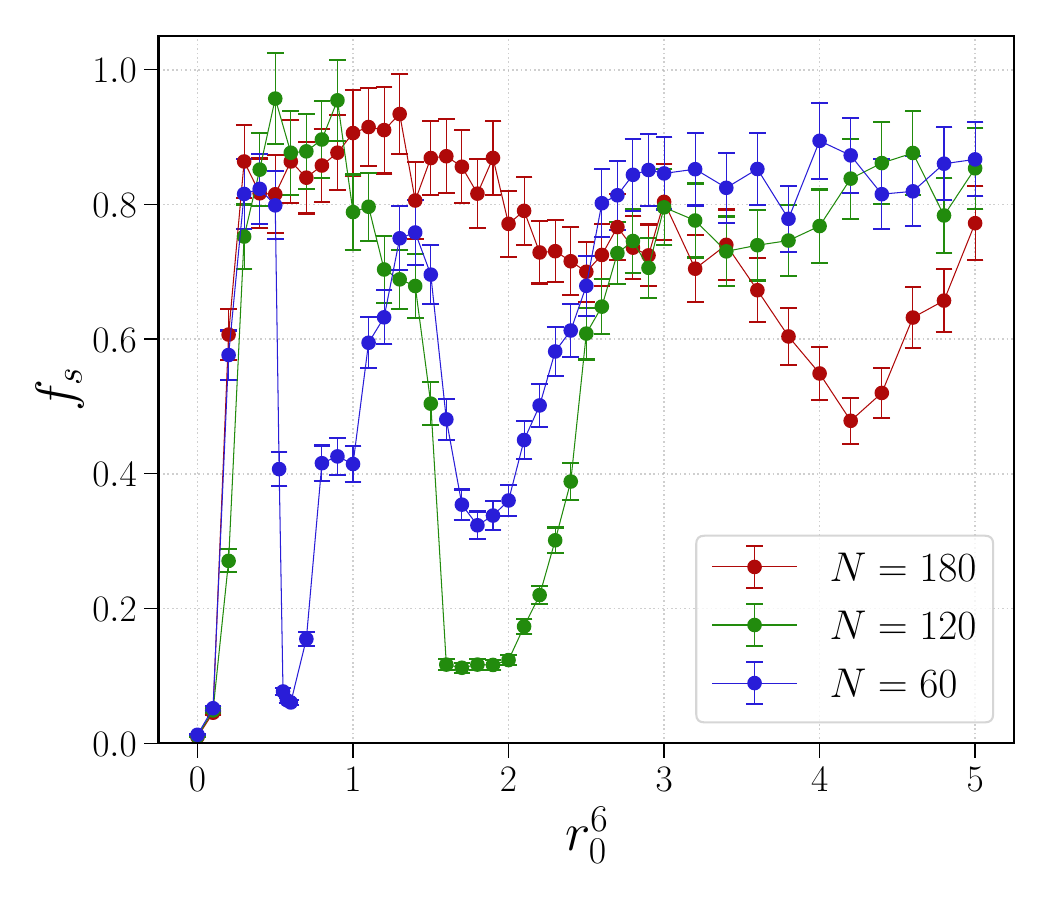}
\caption{Superfluid fraction for $\lambda=0.3$ as a function of system size (in the legend).}
\label{fig12}
\end{figure}
\begin{figure*}[t]
\includegraphics[width=\linewidth]{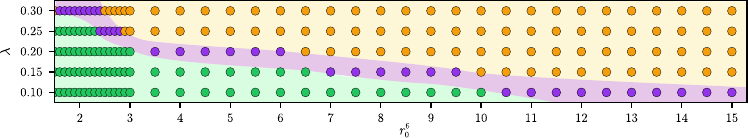}
\caption{$N=60$ bosons: Phase diagram on a much wider range of $r_0^6$ than considered in Fig.~\ref{fig5} or \ref{fig11}. Same color code as in Fig.~\ref{fig5}.}
\label{fig13}
\end{figure*}
For $N=120$ and 180 our exploration of the phase diagram has only been partial, i.e., limited to a narrow interval of $\lambda$ values (from 0.2 to 0.3).
Many of the remarks made for $N=60$ are still valid for the largest sizes.
For example, we see in Fig.~\ref{fig10} that the $r_0^6$ evolution of the integrated radial density is similar to that in Fig.~\ref{fig3}, though, with more bosons, the number of concentric layers in the hard-core case is higher (four for $N=180$).

In Fig.~\ref{fig11} we compare the phase diagrams of the three samples, while plotting in Fig.~\ref{fig12} the superfluid fractions at $\lambda=0.3$.
Like we did before, the phases have been identified by cross-checking information mainly from three sources: superfluid fraction, cycle-length distribution, and visual inspection.
To resolve ambiguities, we have also looked at the RDF and SSF.
The first thing to notice is that the structure changes with increasing $r_0^6$ in a similar way for all sizes:
from multilayer to single layer, from normal to superfluid, and from solid-like to cluster-solid-like.
The supersolids for $N=120$ and $r_0^6\simeq 2$, as well as those for $N=180$ and $r_0^6\simeq 4$ have 12 clusters (lying at the vertices of an icosahedron, see the appendix).
A major difference between the sizes concerns the values of $r_0$ where stable supersolids occur.
Admittedly, requiring the same particle density for all sizes is not sufficient to make the phase behaviors (roughly) coincident.
Perhaps, this happens due to the many variables involved in this comparison:
in the supersolid states, the clusters for different $N$ may occur in different numbers, have different radii, and also different populations.
For example, despite the structure of the supersolid states observed for $\lambda=0.3$ being identical for $N=120$ and $N=180$ (i.e., icosahedral), these states occur for largely different strengths of the interparticle repulsion.

\begin{figure}[t]
\includegraphics[width=\linewidth]{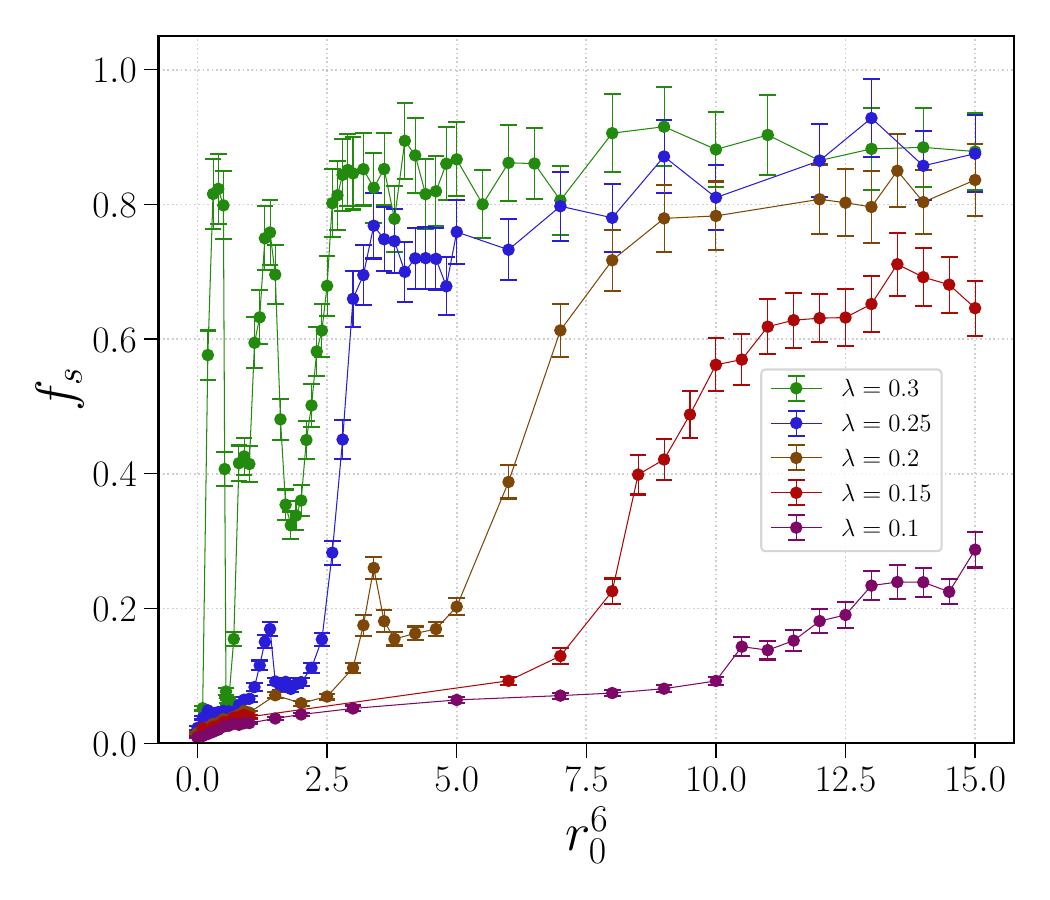}
\caption{$N=60$ bosons: Superfluid fraction for a few values of $\lambda$ (in the legend), plotted as a function of $r_0^6$.}
\label{fig14}
\end{figure}

Finally, we have extended our investigation of the $N=60$ system to values of $r_0^6$ larger than 5 and $\lambda$ smaller than 0.2, to have confirmation that the supersolid phase exists for all $r_0$.
We have thus found other supersolids, see the enlarged phase diagram in Fig.~\ref{fig13} and the superfluid fraction in Fig.~\ref{fig14}, ultimately suggesting the existence of an unbounded region of supersolidity, squeezed between the superfluid region and the cluster-solid region.
Eventually, for extremely large $r_0$ the system must become superfluid (at $T=1$) at practically all $\lambda$.
We conjecture that the behavior is similar for $N=120$ and 180, though the region of supersolidity is probably shifted to much larger values of $r_0^6$.

\section{Conclusions}

Ultracold bosons in a broad shell trap are an unrivaled training ground for many-body physics.
The trap provides a curved confining field which interferes with the natural tendency of atoms to crystallize into a cubic structure.
In small systems (with few tens/hundreds of atoms) the effects of curvature are stronger, and this causes the stabilization of layered solids and supersolids with unprecedented structure~\cite{ciardi2025}.
Two opposite scenarios deserve our attention:
Helium-like atoms, characterized by a harsh mutual repulsion, and Rydberg-dressed atoms, which feature a penetrable core.
While spherical layering of atoms is well known in superfluid Helium, where ``snowballs'' are formed around an impurity~\cite{Buzzacchi2001}, soft-core atoms must first aggregate into clusters before they can give rise to a layered solid.

In this paper we examine the behavior emerging at low temperature when ordinary, Helium-like atoms are gradually ``morphed'' into penetrable atoms, under the influence of a fixed shell-trap potential.
The question is how crystalline order will turn into cluster-solid order, and how big is the chance of observing curved supersolids at the border between superfluid states (at low boson masses) and solidlike states (at high masses).

Using path integral Monte Carlo simulations we have mapped the full phase diagram of bosons with core repulsion ranging from hard to extremely soft (we have set the temperature to a low value and consider three sizes: $N=60,120$, and 180).
We find a rich behavior with solid, supersolid, and superfluid phases, as a function of core softness and boson mass.
Supersolidity is mainly seen in the value of the superfluid fraction (neither low nor high), as well as in the decay mode of the distribution of cycle lengths (close to exponential).
As expected, supersolid order only appears in the presence of clusters.
In our case study, supersolids are only observed in one-layer samples, but this can be changed by properly tuning the parameters of the trap.
The number of clusters in the (super)solids is strongly dependent on the core softness:
it reduces progressively as the particle core becomes increasingly softer.
In particular, we observe supersolids with 12, 6, and 4 clusters, respectively lying at the vertices of an icosahedron, octahedron, and tetrahedron.
The spherical structure factor bears a clear signature of the symmetries of a solidlike arrangement of clusters.

To our knowledge, gases of Rydberg atoms are not currently sufficiently stable to represent a viable, direct experimental realization of this system. Similar physics, however, could be investigated with magnetic atoms, as already done for superfluidity in three-dimensional flat space. 
From previous work, we know that the transition from quantum hard spheres to a strongly dipolar interaction would lead to the formation of a ring supersolid on the equator of the sphere~\cite{ciardi2024}. Our present work indicates that the formation of multiple rings is also possible.

Experimental implementation aside, the model introduced here is relatively simple but captures the whole range of bosonic physics from condensation to supersolidity and crystallization. As such, it should be well positioned as a playground for future studies into thermodynamic and dynamical properties of these phases (such as their response to an external angular velocity field) and how they change across the transitions, especially in the transition from a single-particle crystal to a cluster supersolid.

\begin{acknowledgments}
This work was supported by the European Union through the Next Generation EU funds through the Italian MUR National Recovery and Resilience Plan, Mission 4 Component 2 - Investment 1.4 - National Center for HPC, Big Data and Quantum Computing (CUP B83C22002830001). 
F.~C. acknowledges financial support from PNRR MUR Project No. PE0000023-NQSTI.
M.~C. acknowledges funding from  the Austrian Science Fund (FWF) through the cluster of excellence quantA (Grant No.\ 10.55776/COE1), and from the European Union’s Horizon Europe research and innovation program under the Marie Sklodowska-Curie Grant Agreement No. 101205948 (Q-UDiM).
\end{acknowledgments}

\bibliography{mybose.bib}

\appendix
\section{Spherical structure factor}
\setcounter{equation}{0}
\setcounter{figure}{0}
\renewcommand{\thefigure}{A\arabic{figure}}
\renewcommand{\theequation}{A.\arabic{equation}}

When particles are confined on the surface of a sphere, the expressions of the one-body density function and the RDF have to be adapted to the geometry of the new space.
While the former function indicates where particles are more likely to be found on the sphere, the RDF measures their spatial correlations, telling how many particles are found, on average, at a given (chord) distance from a reference particle.
The spherical structure factor (SSF) conveys a different type of information, as it reveals the symmetries present in the statistical distribution of particle positions.
In this appendix we expand on the relationships that exist between these three quantities.

The running density function is
\begin{equation}
\hat n({\bf r})=\frac{1}{R^2}\delta(r-R)\sum_{i=1}^N\delta^2(\Omega-\Omega_i)\,,
\label{a-1}
\end{equation}
where $\Omega$ is a notation alternative to $\hat{\bf r}$, the unit vector expressing the position of a point particle through its spherical angles $\theta$ and $\phi$.
Clearly, $\hat n({\bf r})$ is also a function of the system configuration.
Its integral over the whole space gives back the total number of particles (in that configuration):
\begin{equation}
\int{\rm d}^2\Omega\,{\rm d}r\,r^2\hat n({\bf r})=\int{\rm d}^2\Omega\sum_{i=1}^N\delta^2(\Omega-\Omega_i)=N\,.
\label{a-2}
\end{equation}
At equilibrium, the (canonical) average of the running density is the one-body density:
\begin{equation}
n({\bf r})=\left\langle\hat n({\bf r})\right\rangle\,,
\label{a-3}
\end{equation}
also normalized to $N$.

We introduce a special symbol, $\rho(\Omega)$, for the angular part of the one-body density:
\begin{equation}
\rho(\Omega)=\left\langle\sum_{i=1}^N\delta^2(\Omega-\Omega_i)\right\rangle\,.
\label{a-4}
\end{equation}
Expanding it on the spherical-harmonics basis,
\begin{equation}
\rho(\Omega)=\sum_{l=0}^\infty\sum_{m=-l}^l\rho_{l,m}Y_l^m(\Omega)\,,
\label{a-5}
\end{equation}
the Fourier coefficients are given by
\begin{eqnarray}
\rho_{l,m}&=&\int{\rm d}^2\Omega\,Y_l^{m\ast}(\Omega)\left\langle\sum_{i=1}^N\delta^2(\Omega-\Omega_i)\right\rangle
\nonumber \\
&=&\left\langle\sum_{i=1}^N\int{\rm d}^2\Omega\,Y_l^{m\ast}(\Omega)\delta^2(\Omega-\Omega_i)\right\rangle
\nonumber \\
&=&\left\langle\sum_{i=1}^NY_l^{m\ast}(\Omega_i)\right\rangle\,.
\label{a-6}
\end{eqnarray}

Consider a specific form of $\rho(\Omega)$, say a Gaussian centered at a particular point $R\hat{\bf r}_1$ of the spherical surface, representing e.g. an isolated cluster of particles on the sphere.
Then,
\begin{equation}
\rho(\Omega)\propto e^{-\frac{(R\hat{\bf r}-R\hat{\bf r}_1)^2}{2\sigma^2}}\propto e^{\gamma\hat{\bf r}\cdot\hat{\bf r}_1}
\label{a-7}
\end{equation}
with $\gamma=R^2/\sigma^2$.
Now assume that particles are equally distributed among $M$ clusters of Gaussian shape, centered at $R\hat{\bf r}_k$.
Since
\begin{equation}
\int{\rm d}^2\Omega\,e^{\gamma\hat{\bf r}\cdot\hat{\bf r}_1}=2\pi\int_0^\pi
{\rm d}\theta\,\sin\theta\, e^{\gamma\cos\theta}=\frac{4\pi}{\gamma}\sinh\gamma\,,
\label{a-8}
\end{equation}
the one-body density of $M$ clusters reads:
\begin{equation}
\rho(\Omega)=\frac{N}{M}\frac{\gamma}{4\pi\sinh\gamma}\sum_{k=1}^Me^{\gamma\hat{\bf r}\cdot\hat{\bf r}_k}\,.
\label{a-9}
\end{equation}
For this angular density, the Fourier coefficients are
\begin{equation}
\rho_{l,m}=\frac{N}{M}\frac{\gamma}{4\pi\sinh\gamma}\sum_{k=1}^M\int{\rm d}^2\Omega\,Y_l^{m\ast}(\Omega)e^{\gamma\hat{\bf r}\cdot\hat{\bf r}_k}\,.
\label{a-10}
\end{equation}
Using the Funk-Hecke formula~\cite{prestipino2019a}, we get:
\begin{equation}
\int{\rm d}^2\Omega\,Y_l^{m\ast}(\Omega)e^{\gamma\hat{\bf r}\cdot\hat{\bf r}_k}=A_lY_l^{m\ast}(\hat{\bf r}_k)
\label{a-11}
\end{equation}
with
\begin{eqnarray}
A_l&=&\int{\rm d}^2\Omega\,e^{\gamma\cos\theta}P_l(\cos\theta)
\nonumber \\
&=&2\pi\int_{-1}^1{\rm d}t\,e^{\gamma t}P_l(t)=4\pi i_l(\gamma)\,,
\label{a-12}
\end{eqnarray}
where $P_l$ is a Legendre polynomial and $i_l$ is a modified Bessel function of the first kind.
In conclusion,
\begin{equation}
\rho_{l,m}=K_l(\gamma)\sum_kY_l^{m\ast}(\hat{\bf r}_k)
\label{a-13}
\end{equation}
with
\begin{equation}
K_l(\gamma)=\frac{N}{M}\frac{\gamma i_l(\gamma)}{\sinh\gamma}\,.
\label{a-14}
\end{equation}
Then, using the addition theorem of spherical harmonics, a particularly compact expression follows for the power spectrum:
\begin{equation}
Q_l^2\equiv\frac{4\pi}{2l+1}\sum_m\left|\rho_{l,m}\right|^2=K_l(\gamma)^2\sum_{k,k'}P_l(\hat{\bf r}_k\cdot\hat{\bf r}_{k'})\,.
\label{a-15}
\end{equation}
The SSF is precisely defined as~\cite{lovsdorfer2019spherical}
\begin{equation}
S_l=\frac{Q_l^2}{N}\,.
\label{a-16}
\end{equation}

\begin{figure}[t]
\includegraphics[width=\linewidth]{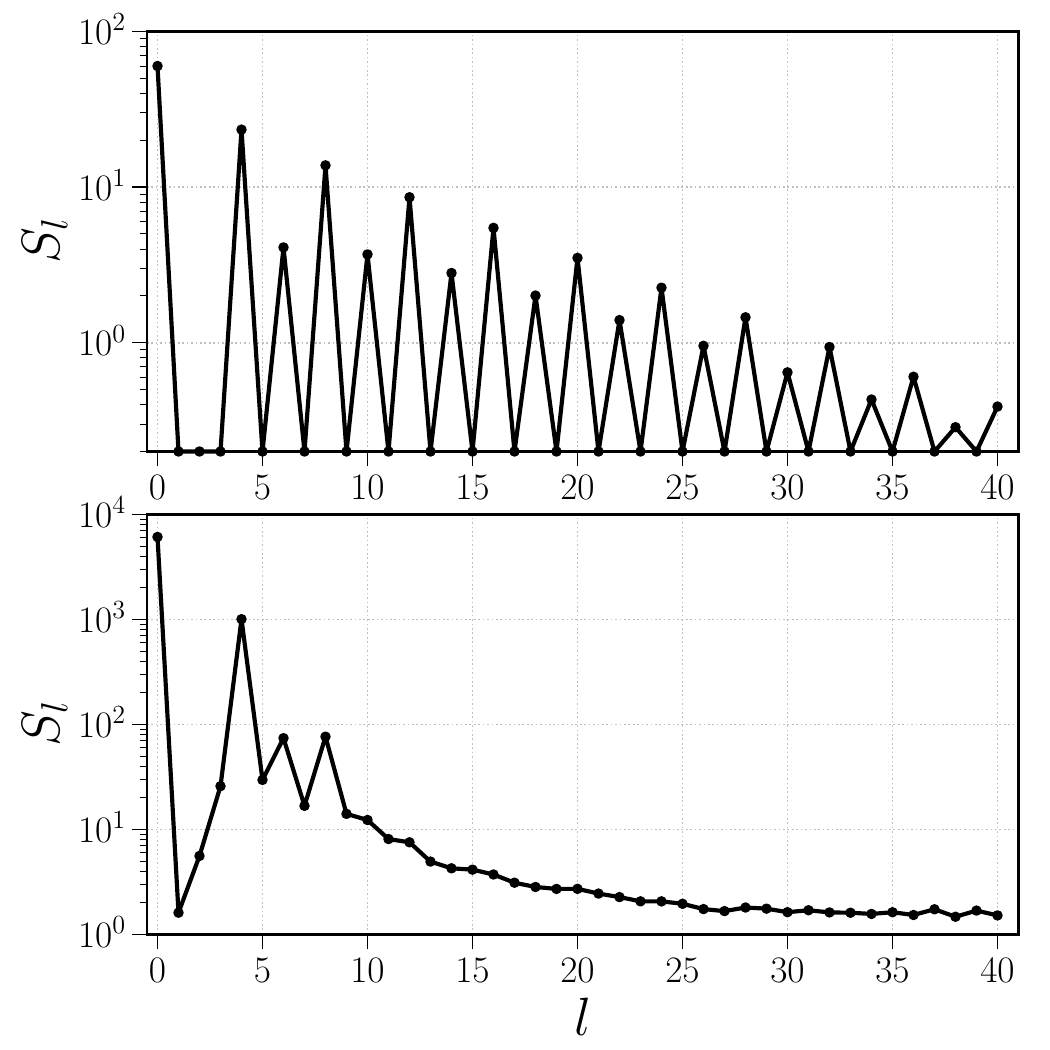}
\caption{$N=60$ bosons. Top: SSF for 6 Gaussian clusters at the vertices of an octahedron.
Notice that, for the sake of visualization, the zero values of $S_l$ at $1,2,3,5,7,9,\ldots$ have been changed to $0.2$.
Bottom: Numerical SSF for $r_0^6=1.8$ and $\lambda=0.3$ (supersolid state).}
\label{figa1}
\end{figure}

\begin{figure}[t]
\includegraphics[width=\linewidth]{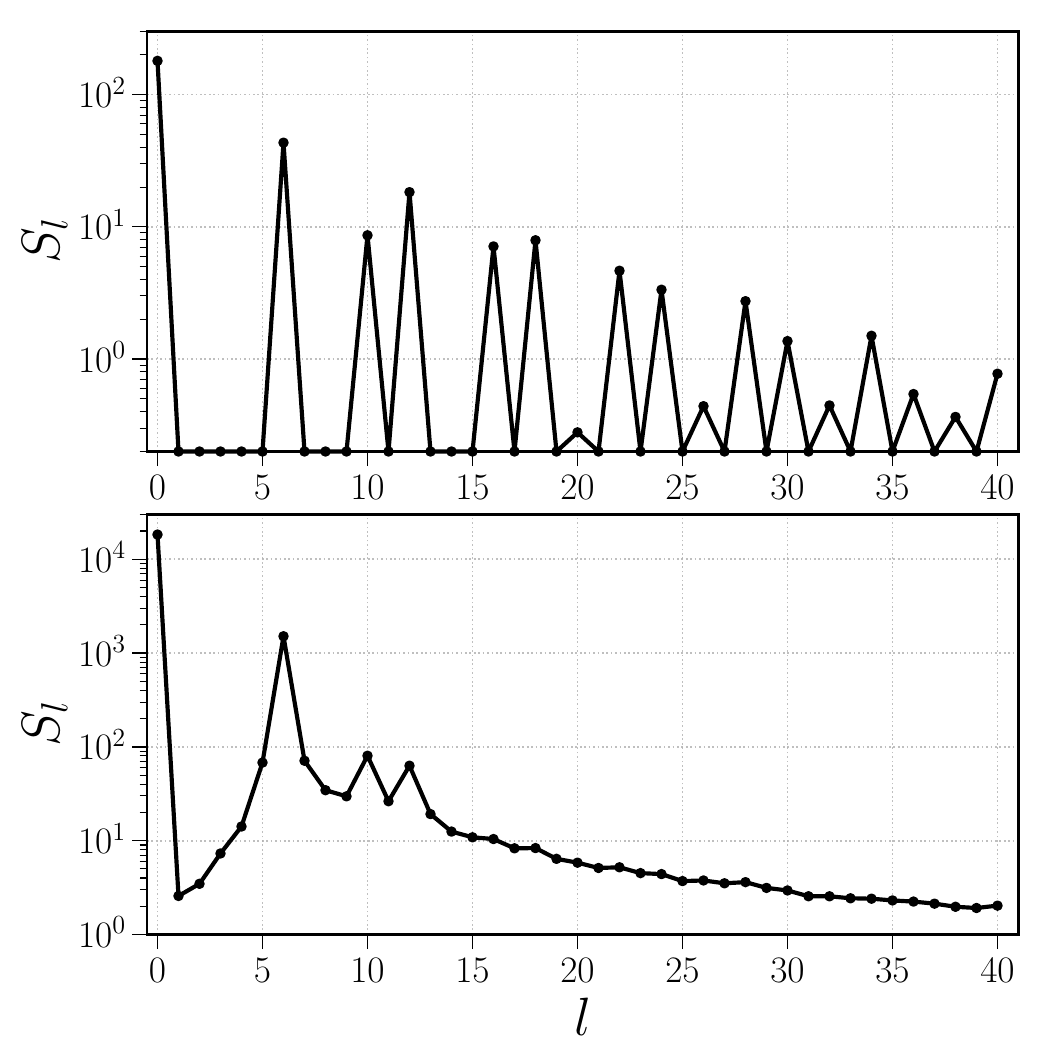}
\caption{$N=180$ bosons. Top: SSF for 12 Gaussian clusters at the vertices of an icosahedron.
Notice that, for the sake of visualization, the zero values of $S_l$ at $1,2,3,4,5,7,\ldots$ have been changed to $0.2$.
Bottom: Numerical SSF for $r_0^6=4.2$ and $\lambda=0.3$ (supersolid state). }
\label{figa2}
\end{figure}

Calling $\delta\hat\rho(\hat{\bf r})=\hat\rho(\hat{\bf r})-\langle\hat\rho(\hat{\bf r})\rangle$ the density fluctuation, we can relate the two-point correlation function
\begin{equation}
G(\hat{\bf r},\hat{\bf r}')=\langle\delta\hat\rho(\hat{\bf r})\delta\hat\rho(\hat{\bf r}')\rangle
\label{a-17}
\end{equation}
of a fluid system of particles on the sphere to the SSF as follows~\cite{lovsdorfer2019spherical}.
Let us introduce the two-body density function,
\begin{equation}
\rho_2(\Omega,\Omega')=2\left\langle\sum_{i<j}\delta^2(\Omega-\Omega_i)\delta^2(\Omega'-\Omega_j)\right\rangle\,,
\label{a-18}
\end{equation}
and the total correlation function,
\begin{equation}
h(\Omega,\Omega')=\frac{\rho_2(\Omega,\Omega')}{\rho(\Omega)\rho(\Omega')}-1\,.
\label{a-19}
\end{equation}
Then, for a homogeneous system ($\rho(\Omega)=N/(4\pi)\equiv\rho$) we find:
\begin{equation}
\rho^2h(\Omega,\Omega')=\langle\hat\rho(\Omega)\hat\rho(\Omega')\rangle-\rho^2-\rho\,\delta^2(\Omega-\Omega')\,.
\label{a-20}
\end{equation}
In practice, $h$ depends on the angular distance between $\Omega$ and $\Omega'$, that is $\hat{\bf r}\cdot\hat{\bf r}'$; therefore,
\begin{equation}
\int{\rm d}^2\Omega\,h(\hat{\bf r}\cdot\hat{\bf r}')Y_l^{m\ast}(\Omega)=h_lY_l^{m\ast}(\hat{\bf r}')
\label{a-21}
\end{equation}
with
\begin{equation}
h_l=\int{\rm d}^2\Omega\,h(\hat{\bf r}\cdot\hat{\bf z})P_l(\cos\theta)=2\pi\int_{-1}^1{\rm d}t\,h(t)P_l(t)\,.
\label{a-22}
\end{equation}
Multiplying both sides of Eq.~(\ref{a-21}) by $Y_{l'}^{m'}(\hat{\bf r}')$ and integrating on $\hat{\bf r}'$, we get:
\begin{equation}
\int{\rm d}^2\Omega\,{\rm d}^2\Omega'\,h(\hat{\bf r}\cdot\hat{\bf r}')Y_l^{m\ast}(\Omega)Y_{l'}^{m'}(\Omega')=h_l\delta_{l,l'}\delta_{m,m'}\,.
\label{a-23}
\end{equation}
Substituting Eq.~(\ref{a-20}) into (\ref{a-23}) we obtain (for $l=l'$ and $m=m'$):
\begin{equation}
\rho(1+\rho h_l)=\left|\rho_{l,m}\right|^2-4\pi\rho^2\delta_{l,0}\delta_{m,0}\,,
\label{a-24}
\end{equation}
which finally yields
\begin{equation}
S_l=1+\rho h_l+N\delta_{l,0}\,.
\label{a-25}
\end{equation}

\begin{figure}[b]
\includegraphics[width=\linewidth]{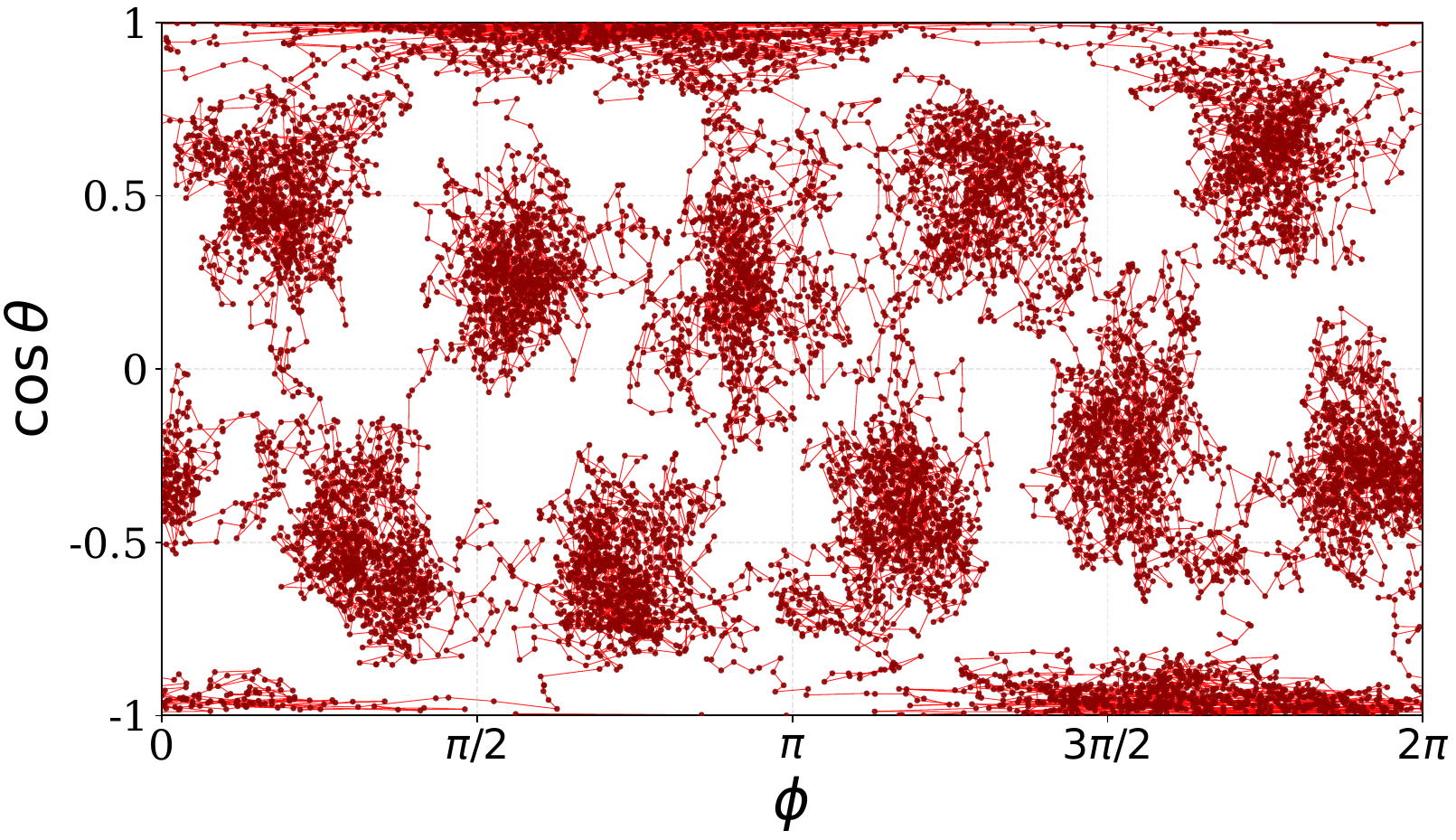}
\caption{$N=180$ bosons:
Typical bead configuration for $r_0^6=4.2$ and $\lambda=0.3$ (projection map).}
\label{figa3}
\end{figure}

We compute the SSF in Eq.~(\ref{a-16}) by averaging over 50 bead configurations taken at regular intervals from the production run of our simulation (Fig.~\ref{fig9} is obtained this way).
As commented in Section III, the locations of the SSF peaks are an imprint of the symmetries inherent in the cluster distribution.
To provide a demonstration, we have computed the power spectrum (\ref{a-15}) by assuming Gaussian clusters centered at the vertices of an octahedron or icosahedron, with $\gamma=0.01$ (an arbitrary small value).
The resulting profiles, to which an exponential damping is applied in order to simulate the presence of noise, are compared in Figs.~\ref{figa1} and \ref{figa2} with the SSF of 60 bosons ($r_0^6=1.8$ and $\lambda=0.3$) and 180 bosons ($r_0^6=4.2$ and $\lambda=0.3$), respectively.
As is clear from the figures, the position and relative height of the first few SSF peaks are exactly reproduced by Eq.~(\ref{a-15}), which certifies the existence of the symmetry.
For $N=180$, this is visually confirmed in Fig.~\ref{figa3}, where we report the projection map of a typical bead configuration for $r_0^6=4.2$ and $\lambda=0.3$.
This map is completely congruent with the red spots in Fig.~3a of Ref.~\onlinecite{cinti2026}, which also correspond to an icosahedral arrangement of clusters.
\end{document}